\documentclass{article}
\usepackage{caption} 

\usepackage{PRIMEarxiv}

\usepackage[utf8]{inputenc} 
\usepackage[T1]{fontenc}    
\usepackage{hyperref}       
\usepackage{url}            
\usepackage{booktabs}       
\usepackage{amsfonts}       
\usepackage{nicefrac}       
\usepackage{microtype}      
\usepackage{lipsum}
\usepackage{appendix}
\usepackage[most]{tcolorbox}
\usepackage{caption}
\usepackage{tcolorbox}
\usepackage{array}
\usepackage{amsmath}
\usepackage{svg}
\usepackage{fancyhdr}       
\usepackage{graphicx}       
\graphicspath{{media/}}     

\title{BioAtt: Anatomical Prior Driven Low-Dose CT Denoising}

\author{
  Namhun Kim\thanks{These authors contributed equally to this work.},  
  Ui Hyun Cho\footnotemark[1] \\  
  Yonsei University \\  
  Seoul, South Korea \\  
  \texttt{\{ksouth0413, cuihyun12\}@yonsei.ac.kr}  
}

\begin{document}
\maketitle

\begin{abstract}
Deep-learning based denoising methods have significantly improved Low-Dose CT (LDCT) image quality. However, existing models often over-smooth important anatomical details due to their purely data-driven attention mechanisms. To address this challenge, we propose a novel LDCT denoising framework, \textbf{BioAtt}. The key innovation lies in attending anatomical prior distributions extracted from the pretrained vision–language model BiomedCLIP. These priors guide the denoising model to focus on anatomically relevant regions to suppress noise while preserving clinically relevant structures. We highlight three main contributions: BioAtt outperforms baseline and attention-based models in SSIM, PSNR, and RMSE across multiple anatomical regions. The framework introduces a new architectural paradigm by embedding anatomic priors directly into spatial attention. Finally, BioAtt attention maps provide visual confirmation that the improvements stem from anatomical guidance rather than increased model complexity.
\end{abstract}


\section{Introduction}

Low-Dose Computed Tomography (LDCT) minimizes patient radiation exposure at the cost of obscured diagnostic details from increased noises and artifacts. Deep-learning denoising methods address this issue by learning a direct mapping from noisy images to their denoised counterparts \cite{Pandey2019RecentDL, Zhang2023ResearchPO, Elshafai2023TraditionalAD, Nazir2024RecentDI, zhang2024review}. For instance,  Residual Encoder-Decoder Convolutional Neural Network (\textbf{RED-CNN}) \cite{chen2017lowdose}, \textit{Yuan et al.} \cite{Yuan2019LowdoseCC}, and DRCNN \cite{Jiang2019ScatterCO} combine autoencoder frameworks with residual connections to suppress noise while maintaining structural information. These early CNN-based approaches \cite{Yang2017CTID, GholizadehAnsari2019DeepLF, Ataei2020CascadedCN} exhibit competitive denoising performance compared to traditional techniques, such as Gaussian Filtering \cite{geraldo2016low}, Weighted Nuclear Norm Minimization \cite{huang2021weighted}, Sinogram Domain Filtering \cite{Yu2008SinogramSW}, and BM3D Methods \cite{zhou2020supervised}. However, these data-driven networks tend to over-smooth important anatomical details \cite{Zhang2023StructurepreservingLC, Chen2023ASCONAS}. Recent studies explore attention mechanisms and domain-specific priors to preserve anatomical details into the denoising process \cite{Yang2017LowDoseCI, Jia2023ADC, Zhu2024SAMAttAP, Zhang2024ANN}.

Attention modules play a critical role in improving denoising performance of LDCT models. Squeeze-and-Excitation (\textbf{SE}) network \cite{hu2018squeeze} introduced channel-wise attention to adaptively recalibrate feature responses. Sequentially, Convolutional Block Attention Module (\textbf{CBAM}) \cite{woo2018cbam} applied channel and spatial attention to intermediate feature representations. In the field of LDCT denoising, \textbf{SAM-Att} \cite{Zhu2024SAMAttAP} incorporated a residual attention mechanism into CNN architectures by combining upstream low-rank fine-tuning with downstream attention modules. \textbf{DEMACNN} \cite{Jia2023ADC} integrated edge extraction from both input images and intermediate feature maps with multi-scale attention and a compound loss function. This architecture outperformed RED-CNN, transformer-based models, and other state-of-the-art methods across various LDCT image datasets. Similarly, \textit{Zhang et al.} \cite{Zhang2024ANN} proposed a two-branch multi-scale residual attention network that fuses shallow and deep features to enhance texture and structure recovery in low-dose CT images. These attention-based strategies show strong potential at reducing noise while preserving intricate anatomical details.

In parallel, researchers began to integrate anatomical priors and semantic information to enhance structural preservation of LDCT images. \textit{Huang et al.} \cite{Huang2020ConsideringAP} introduced anatomical site labels as training attributes into a Wasserstein GAN for LDCT enhancement. This method enabled model to adaptively adjust to region-specific features. \textbf{ASCON} \cite{Chen2023ASCONAS} utilized inherent anatomical semantics to guide LDCT denoising through a novel supervised contrastive learning framework that enforces anatomical consistency between output and ground truth. The network integrated multi-scale contrastive modules with self-attention-based U-Net to enhance interpretability of tissue-specific structural details. \textit{Huang et al.} \cite{Huang2022SegmentationguidedDN} also segmented preliminary organ or tissue segmentation as priors to direct the denoising process toward anatomically relevant regions. While these approaches improve interpretability and image quality, they introduce additional complexity by relying on extra labels or separate segmentation modules.

To address these limitations, we propose \textbf{BioAtt}, a novel LDCT denoising framework that integrates organ-aware spatial attention module \cite{Feng2020MultiDimensionalSA, Marcos2021LowDC} into the RED-CNN architecture using anatomical prior probability distribution \(\mathbf{p}\) derived from BiomedCLIP \cite{zhang2024biomedclip}. Building on RED-CNN’s residual learning strategy, BioAtt embeds a BiomedCLIP-guided attention module within the encoder-decoder pipeline to better preserve anatomical fidelity. Our key modification lies in the organ-aware spatial attention block: \textbf{BioAtt block}, which is embedded within the encoder–decoder pipeline. This block receives intermediate feature vector \(\mathbf{x}\) along with anatomical prior vector \(\mathbf{p}\) to generate reweighted feature vector \(\mathbf{x'}\). Unlike conventional attention mechanisms that rely solely on data-driven feature patterns and lack explicit semantic context \cite{Zhang2023AND, Li2023MultiScaleFF, Zhang2024ANN}, BioAtt introduces a fundamentally new architectural paradigm by embedding anatomical priors directly into its attention mechanism. The model is explicitly informed about the presence of anatomical structures from the semantically meaningful and organ-aware BioAtt block. This not only enhances interpretability but also improves denoising performance by guiding the network’s focus toward clinically relevant regions.

We present three key contributions of BioAtt: (1) BioAtt outperforms not only the baseline RED-CNN but also other attention-based denoising models by achieving higher SSIM, PSNR and lower RMSE across multiple anatomical regions including lungs, liver, and kidneys. (2) BioAtt introduces a novel paradigm by embedding anatomical priors into the network’s attention mechanism derived from a pretrained vision–language model. Our method explicitly incorporates organ-aware spatial attention to yield more semantically richer feature modulation. (3) We verify that the improvement stems from anatomical guidance rather than increased model complexity. Ablation studies confirm that models incorporating BiomedCLIP-based priors consistently outperform counterparts lacking such semantic information. Also, attention maps demonstrate that BioAtt preserves organ-specific focus across layers which offers visual explainability absent in several mechanisms.

\section{Methodology}

\begin{figure*}[h] 
\centering 
\includegraphics[width=1.0\textwidth]{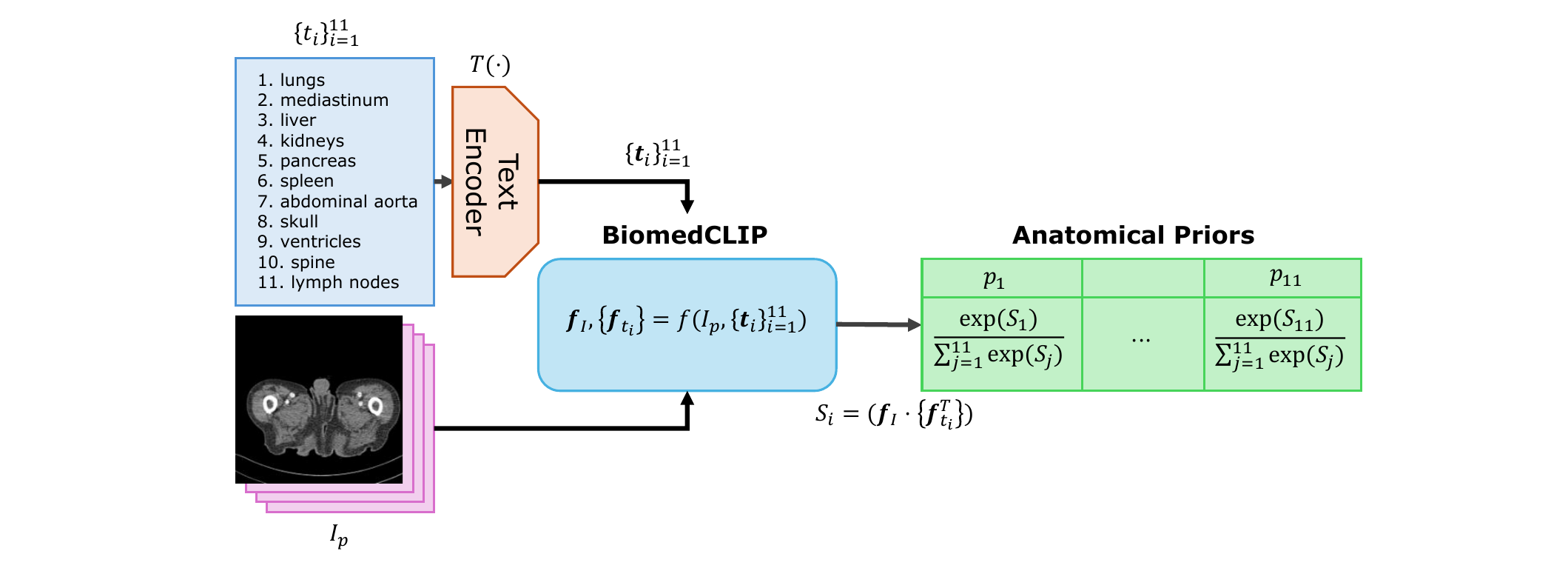} 
\caption{Overview of anatomical prior extraction process. Given a low-dose CT image \(\mathbf{I}_p\), a set of anatomical descriptors \(\{t_i\}_{i=1}^{N}\) is tokenized and encoded using a pretrained text encoder. BiomedCLIP jointly embeds the image and text to compute similarity scores \(S_i\), which are normalized via softmax to yield a probability distribution \(\mathbf{p}\). Each \(p_i\) reflects the estimated likelihood of a specific anatomical structure in the input image. These anatomical priors guide spatial attention in the denoising network.}
\label{fig1} 
\end{figure*}

BioAtt extends the classic RED-CNN encoder–decoder design by integrating organ-aware spatial attention modules that are guided by anatomical priors. Specifically, given an input LDCT image \(I_p \in \mathbb{R}^{1 \times H \times W}\), anatomical priors are first extracted using a pretrained vision–language model (Section~\ref{2_1}). These priors are then incorporated into a spatial attention mechanism (Section~\ref{2_2}), which generates organ-weighted attention maps to modulate intermediate feature representations. Feature maps are refined to emphasize semantically important regions while suppressing background noise and the decoder then reconstructs the denoised output \(\hat{Y} \in \mathbb{R}^{1 \times H \times W}\) using deconvolution layers and residual skip connections (Section~\ref{2_3}).

\subsection{Extracting Anatomical Priors from BiomedCLIP}
\label{2_1}

We utilize the pretrained BiomedCLIP model \(f(\cdot)\) to extract anatomical priors from LDCT image \(\mathbf{I}_p\) (Figure~\ref{fig1}). The objective is to compute a probability distribution \(\mathbf{p}\) over a set of anatomical descriptors\(\{t_i\}_{i=1}^N\), thereby enhancing the denoising process with domain-specific anatomical information. First, the set of anatomical descriptors \(\{t_i\}_{i=1}^N\) is tokenized using a pretrained BertTokenizer \(T(\cdot)\) to yield \(\mathbb{R}^{512}\) sized text embeddings. Then, the BiomedCLIP model \(f(\cdot)\) processes LDCT image \(\mathbf{I}_p\) along with the corresponding tokenized descriptors \(\{\mathbf{t}_i\}\). This results in an image vector \(\mathbf{f}_I\) and a set of text feature vectors \(\{\mathbf{f}_{t_i}\}\) (Equation~\ref{eq1}).

\begin{equation}
\label{eq1}
\begin{aligned}
\mathbf{f}_I, \{\mathbf{f}_{t_i}\} = f(I_p, \{\mathbf{t}_i\}), \quad i = 1, ..., N.
\end{aligned}
\end{equation}

To estimate the relevance of each anatomical descriptor \(t_i\), we compute a similarity score \(S_i\) between the image vector \(\mathbf{f}_I\) and the \(i\)-th text embeddings  \(\mathbf{f}_{t_i}\) using the dot product \((\mathbf{f}_I\cdot\mathbf{f}_{t_i}^T)\). A softmax operation is then applied to yield a probability distribution \(p_i\) (Equation~\ref{eq2}). Thus, \(\mathbf{p} = [p_1, p_2, ... , p_N]^T\) successfully represents the likelihood of the presence of each anatomical structure.

\begin{equation}
\label{eq2}
p_{i} = \frac{\exp(S_i)}{\sum_{j=1}^N \exp(S_j)}, \quad i = 1, ..., N.
\end{equation}

\subsection{Guiding Spatial Attention with Anatomical Priors}
\label{2_2}

\begin{figure*}[h] 
\centering 
\includegraphics[width=1.0\textwidth]{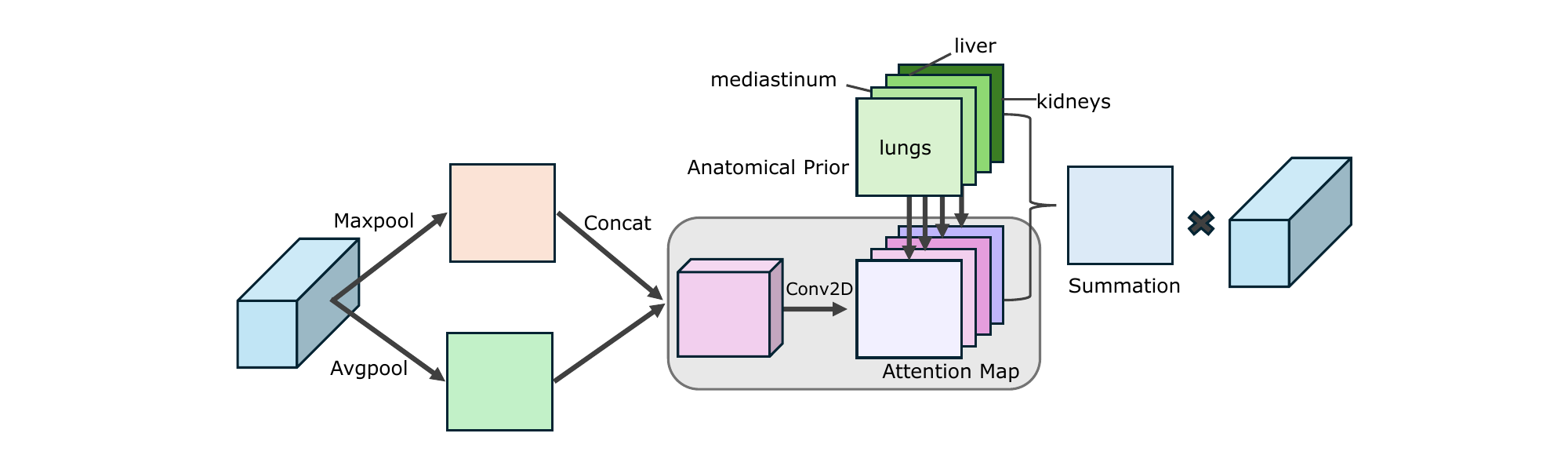} 
\caption{Overview of the organ-aware spatial attention module. The input feature map undergoes both average and max pooling along the channel axis. Descriptors are then concatenated and passed through a convolutional layer. This produces a multi-channel attention map corresponding to \(N\) different anatomical structures. The attention maps are then modulated by anatomical prior probabilities and summed across organs to yield a unified spatial attention map, which is applied back to the original feature map to emphasize clinically relevant regions.}
\label{fig2} 
\end{figure*}

We incorporate anatomical priors into a spatial attention mechanism to reflect feature representation into the denoising process (Figure~\ref{fig2}. By selectively focusing on spatial regions associated with likely anatomical structures, this module preserves clinically relevant details while effectively suppressing noise. Let \(\mathbf{x} \in \mathbb{R}^{B \times C \times H \times W}\) denote the input feature map where \(B\) is the batch size, \(C\) is the number of channels, and \(H\), \(W\) are the feature map dimensions. We compute two complementary spatial descriptors through channel-wise pooling operations (Equation~\ref{eq3}).

\begin{equation}
\label{eq3}
\text{avg\_out} = \frac{1}{C}\sum_{c=1}^C\mathbf{x}_{B,c,H,W} \quad \text{and} \quad \text{max\_out} = \max_{c = 1,2,...,C}\mathbf{x}_{B,c,H,W}
\end{equation}

These operations yield tensors of shape \([B, 1, H, W]\), which are concatenated along the channel dimension to form a combined representation \(\mathbf{C} \in \mathbb{R}^{B \times 2 \times H \times W}\). We then apply a 2D convolutional layer with kernel size \(k = 7\) to produce a set of attention map \(\mathbf{A}\) corresponding to \(N\) anatomical structures. \(\sigma(\cdot)\) denotes the sigmoid activation function. (Equation~\ref{eq4})

\begin{equation}
\label{eq4}
\mathbf{A} = \sigma(\text{Conv2D}(\mathbf{C};\theta)) \in \mathbb{R}^{B \times N \times H \times W}
\end{equation}

Next, we reshape the probability distribution \(\mathbf{p} \in \mathbb{R}^{B \times N}\) to \(\mathbf{p'}\) to align with the spatial dimensions of the attention map \(\mathbf{A}\). We perform element-wise multiplication to produce weighted attention map \(\mathbf{W}\) that modulates the attention maps with anatomical relevance (Equation~\ref{eq5}).

\begin{equation}
\label{eq5}
\mathbf{W} = \mathbf{A} \odot \mathbf{p'} \in  \mathbb{R}^{B \times N \times H \times W}
\end{equation}

Finally, we aggregate the weighted maps across the anatomical dimension to obtain a unified attention map \(\mathbf{A'}\) (Equation~\ref{eq6}). This final attention map is applied to the original feature map in order to make the network focus on semantically meaningful regions during denoising process. \(\mathbf{x'} = \mathbf{x} \odot \mathbf{A'}\).

\begin{equation}
\label{eq6}
\mathbf{A'} = \sum_{n=1}^N W_{[:,n,:,:]} \in  \mathbb{R}^{B \times 1 \times H \times W}
\end{equation}

\subsection{Network Architecture with Organ-Aware Spatial Attention}
\label{2_3}

\begin{figure*}[h] 
\centering 
\includegraphics[width=0.7\textwidth]{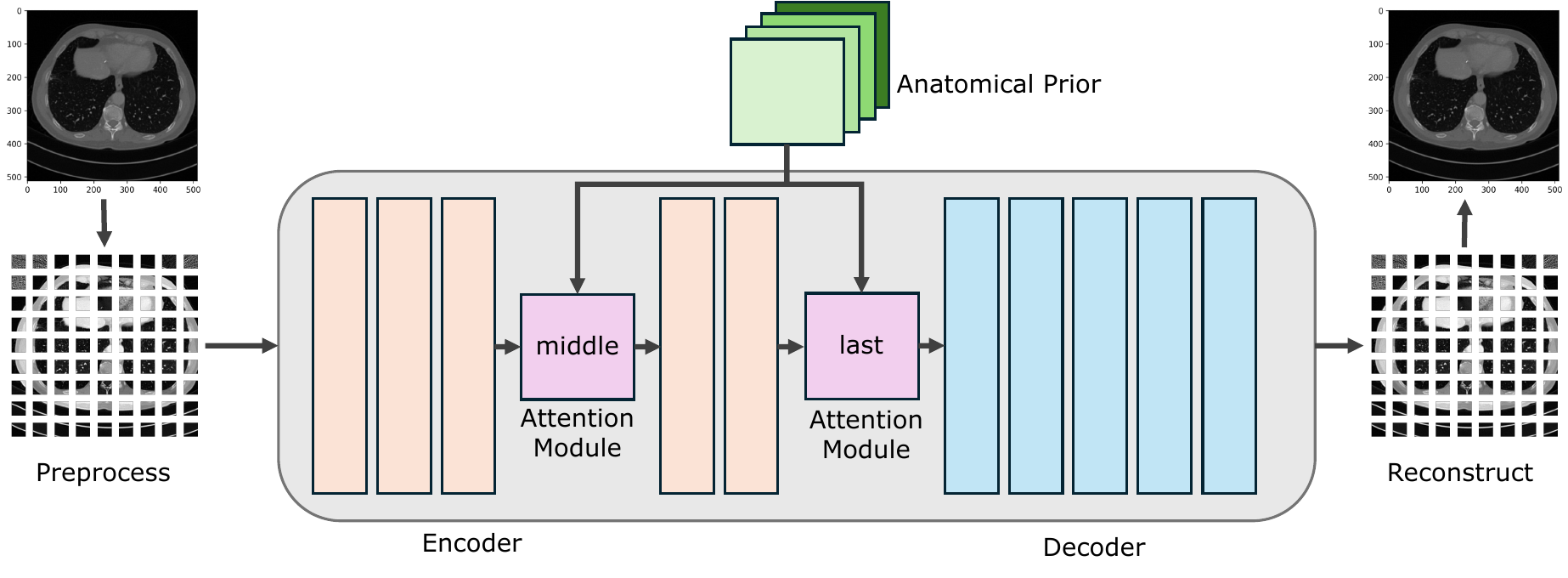} 
\caption{Overall architecture of BioAtt. The input low-dose CT image is first divided into patches. These patches are passed through an encoder composed of convolutional layers with two spatial attention modules guided by anatomical priors. The decoder then reconstructs the denoised image from the refined feature maps to restore the full-resolution CT image.}
\label{fig3} 
\end{figure*}

BioAtt is an encoder–decoder network that integrates anatomical priors into the denoising process through organ-aware spatial attention (Figure~\ref{fig3}). The architecture builds upon the RED-CNN framework by embedding attention mechanisms that utilize anatomical probability maps derived from Section~\ref{2_1}. These priors guide the network to emphasize clinically relevant regions while suppressing noise.

\textbf{Encoder: Feature Extraction with Attention Integration.} The encoder consists of a sequence of convolutional layers for hierarchical feature extraction. \textit{Two spatial attention modules} are strategically embedded within the encoder of RED-CNN, aligning with its intermediate layers. These modules refine two intermediate feature maps (\textit{middle}, \textit{last}) by modulating them with organ-specific attention weights. This mechanism enables the network to selectively enhance features corresponding to key anatomical regions.

\textbf{Decoder: Image Reconstruction through Deconvolution.}
The decoder mirrors the encoder with a series of deconvolution layers that progressively upsample the feature maps to reconstruct the denoised CT image. The network reconstructs the subset that is patchified from the input image to produce the final output.

By embedding anatomically guided spatial attention modules at critical stages within the encoding process, BioAtt enables the network to better preserve anatomical structures while effectively reducing noise. This organ-aware design is thus well-suited for clinical LDCT image enhancement tasks.

\section{Experiments}

\subsection{Datasets and Training}

We utilize the NIH-AAPM-Mayo Clinic Low Dose CT Grand Challenge 2016 (Mayo-2016) dataset~\cite{mccollough2017lowdose} for the image denoising task. Specifically, we designate the \texttt{quarter\_1mm} images for Low-Dose CT (LDCT) data and the \texttt{full\_1mm} images for Normal-Dose CT (NDCT) data. We first apply standardization using a mean of \(-500\) Hounsfield Units (HU) and a standard deviation of 500 HU. This approach aligns well with our experimental objectives, as it effectively preserves the underlying intensity distribution while ensuring numerical stability across different image samples.

\begin{center}
    \includegraphics[width=1.0\textwidth]{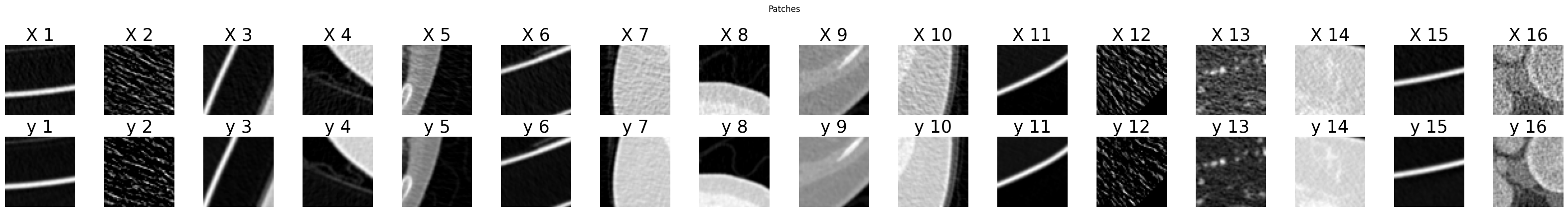}
    \captionof{figure}{Comparison of \(16\) \texttt{quarter\_1mm} and \texttt{full\_1mm} patches preprocessed from Mayo-2016 Dataset.}
    \label{fig9}
\end{center}

Next, we patchify each images to reduce computational complexity while preserving local structural information. Each \(512 \times 512\) images are divided into \(55 \times 55\) patches, yielding a total of 81 patches per image (a \(9 \times 9\) grid). To further augment the training data, we randomly rotate images with a fixed probability. Finally, the dataset is split into \(0.64\) for training, \(0.16\) for validation, and \(0.20\) for testing, with a batch size of \(16\). Consequently, the number of patches per dataset was 57,631 for training, 4,815 for validation, and 6,025 for testing.

Finally, we use the Adam optimizer to train our model, with the Mean Squared Error (MSE) loss function. The initial learning rate is set to \(1 \times 10^{-5}\) and it is reduced by half every 5 epochs, with a minimum threshold of \(1 \times 10^{-10}\). Training ends when no improvement is observed for 7 consecutive evaluations.

\subsection{Experiment 1. Comparative Evaluation of Attention Mechanisms}

\begin{center}
    \includegraphics[width=0.7\textwidth]{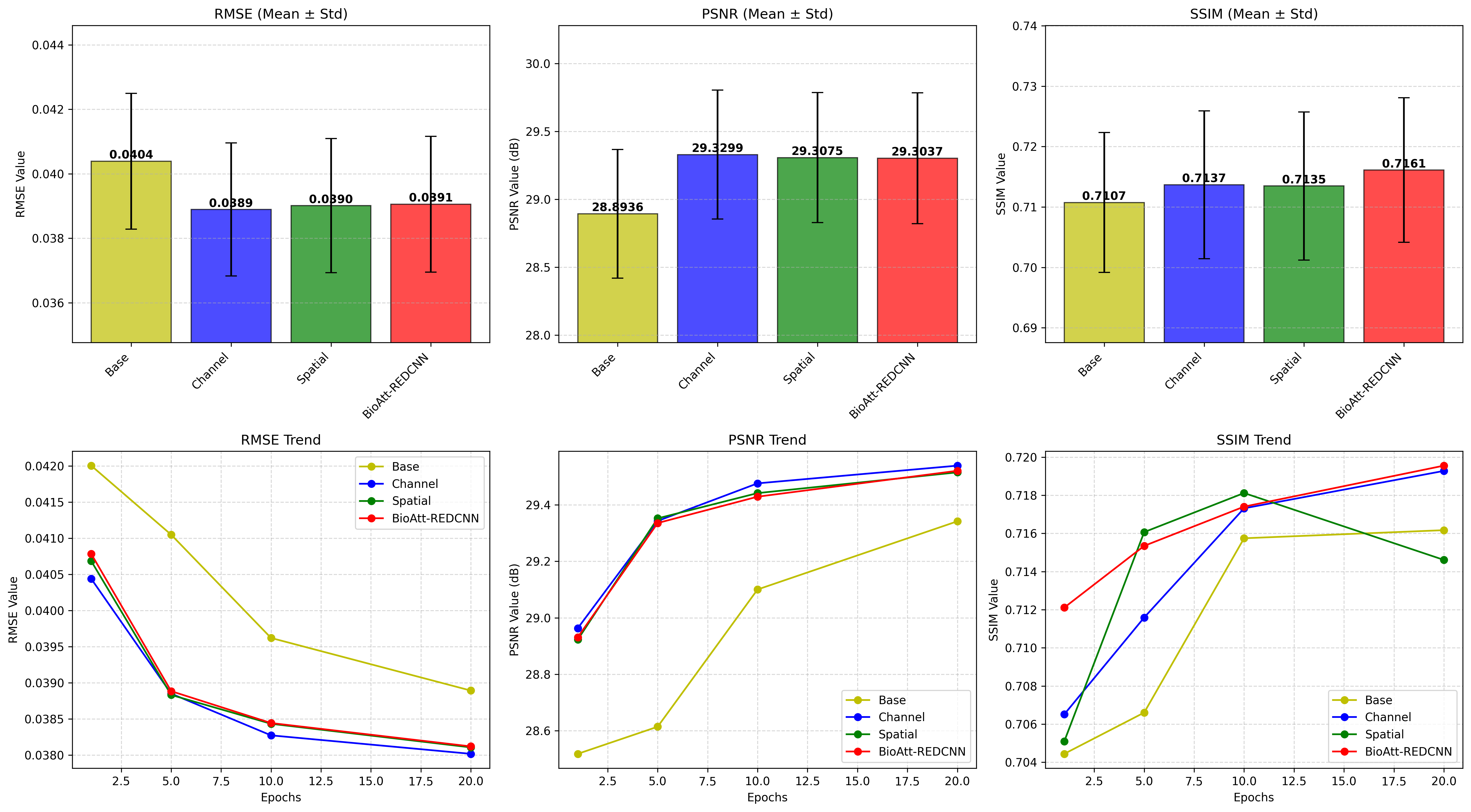}
    \captionof{figure}{Performance comparison of four models: \textit{Base}, \textit{Channel}, \textit{Spatial}, and \textit{BioAtt}. The top row shows evaluation metrics (RMSE, PSNR, and SSIM) with mean and standard deviation across test samples. The bottom row illustrates the trend of each metric over training epochs (1, 5, 10, and 20). While all attention-augmented models outperform the baseline, \textit{BioAtt} consistently achieves higher SSIM and demonstrates stable improvements throughout training.}
    \label{fig4}
\end{center}

We conducted a comparative evaluation between the baseline RED-CNN model (\textit{Base}) with three attention-augmented variants: a channel attention model (\textit{Channel}), a spatial attention model (\textit{Spatial}), and our proposed model (\textit{BioAtt}). 

The channel attention mechanism employs the Squeeze-and-Excitation (SE) block, which adaptively recalibrates channel-wise feature responses \cite{hu2018squeeze}. The spatial attention mechanism utilizes the Convolutional Block Attention Module (CBAM), which applies attention across spatial dimensions to highlight informative regions within the feature vectors \cite{woo2018cbam}. We insert these three separate attention modules after the third and fifth convolutional layers in the encoder. This setup allows the network to better understand feature representations during the intermediate stages, thereby improving the model's ability to capture relevant structures in LDCT denoising.

As shown in the first row of Figure~\ref{fig4}, attention mechanisms consistently improve performance over the baseline model. While the \textit{Channel} model achieves the best results in terms of RMSE and PSNR, \textit{BioAtt} shows the highest SSIM, which indicates better structural preservation.

The second row shows metric trends over training epochs (1, 5, 10, and 20) for the best-performing models. As all models use MSE loss, RMSE decreases and PSNR increases steadily. Interestingly, SSIM exhibits a contrasting trend: \textit{Spatial} shows an initial increase followed by a decline, whereas both \textit{Channel} and \textit{BioAtt} show consistent improvement throughout training. Notably, \textit{BioAtt} exhibits a stable upward SSIM trend, which reflects its ability to preserve semantic structure.

This behavior suggests that while \textit{Spatial} attention tends to focus on local features, \textit{Channel} attention captures more global contextual information. Our \textit{BioAtt} combines spatial attention with anatomical priors to simultaneously emphasize localized anatomical regions and maintain global structural consistency.

\subsection{Experiment 2. Comparative Evaluation of Patching Mechanisms}

The original RED-CNN framework~\cite{chen2017lowdose} employed a patch-based training strategy to improve denoising performance. Patch-based learning offers several benefits in deep learning: it facilitates data augmentation, increases the diversity of training samples, and enables the model to better capture local structural patterns. By focusing on small image regions, the network learns fine-grained features essential for effective LDCT denoising.

In contrast, our proposed BioAtt framework integrates BiomedCLIP to extract global semantic priors from full images before feeding features into the spatial attention module. Since BiomedCLIP operates on full images to retain overall anatomical context, it is important to evaluate whether whole-image training might better complement this global understanding.

To investigate this, we compare two training strategies: (1) The whole-image training: Full $512 \times 512$ images are used directly with a batch size of $1$. (2) The conventional patch-based training: Each $512 \times 512$ CT images are divided into $55 \times 55$ patches and trained with a batch size of $16$.

\begin{center}
    \includegraphics[width=0.7\textwidth]{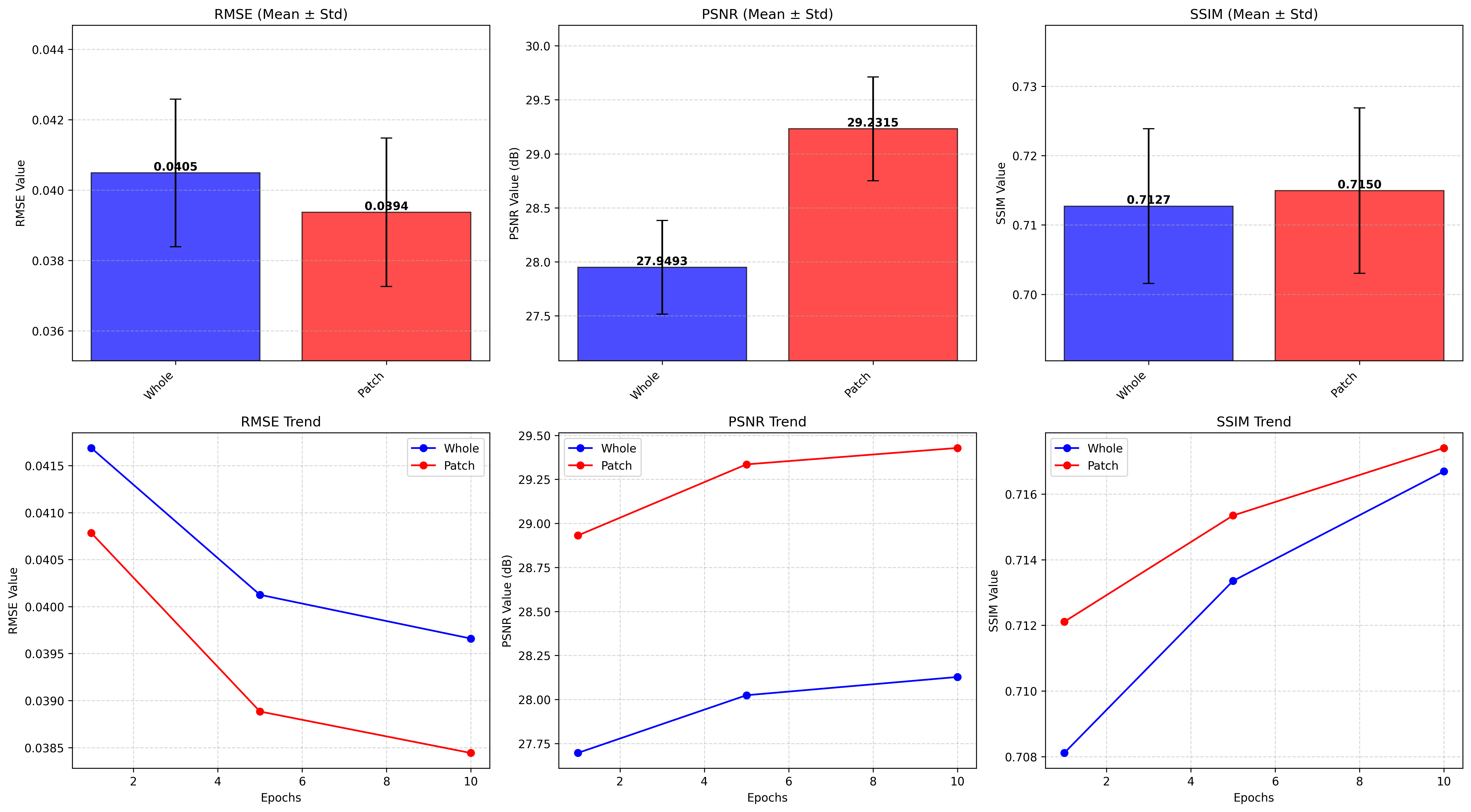}
    \captionof{figure}{Comparison between whole-image and patch-based training strategies across all evaluation metrics. The top row presents the mean and standard deviation of RMSE, PSNR, and SSIM for each strategy. The patch-based approach consistently outperforms the whole-image method. The bottom row shows the progression of these metrics over training epochs. These results emphasize the effectiveness of localized feature learning in LDCT denoising, even when anatomical priors are incorporated.}
    \label{fig5}
\end{center}

Experimental results (Figure~\ref{fig5}) show that the patch-based strategy consistently outperforms whole-image training across RMSE, PSNR, and SSIM. This indicates that while anatomical prior \(\mathbf{p}\) consists global context in its distribution, localized feature extraction through patch-based training remains crucial to capture fine anatomical structures and to enhance LDCT denoising performance.

\subsection{Experiment 3. Comparative Evaluation of Weighting}

\begin{center}
    \includegraphics[width=0.7\textwidth]{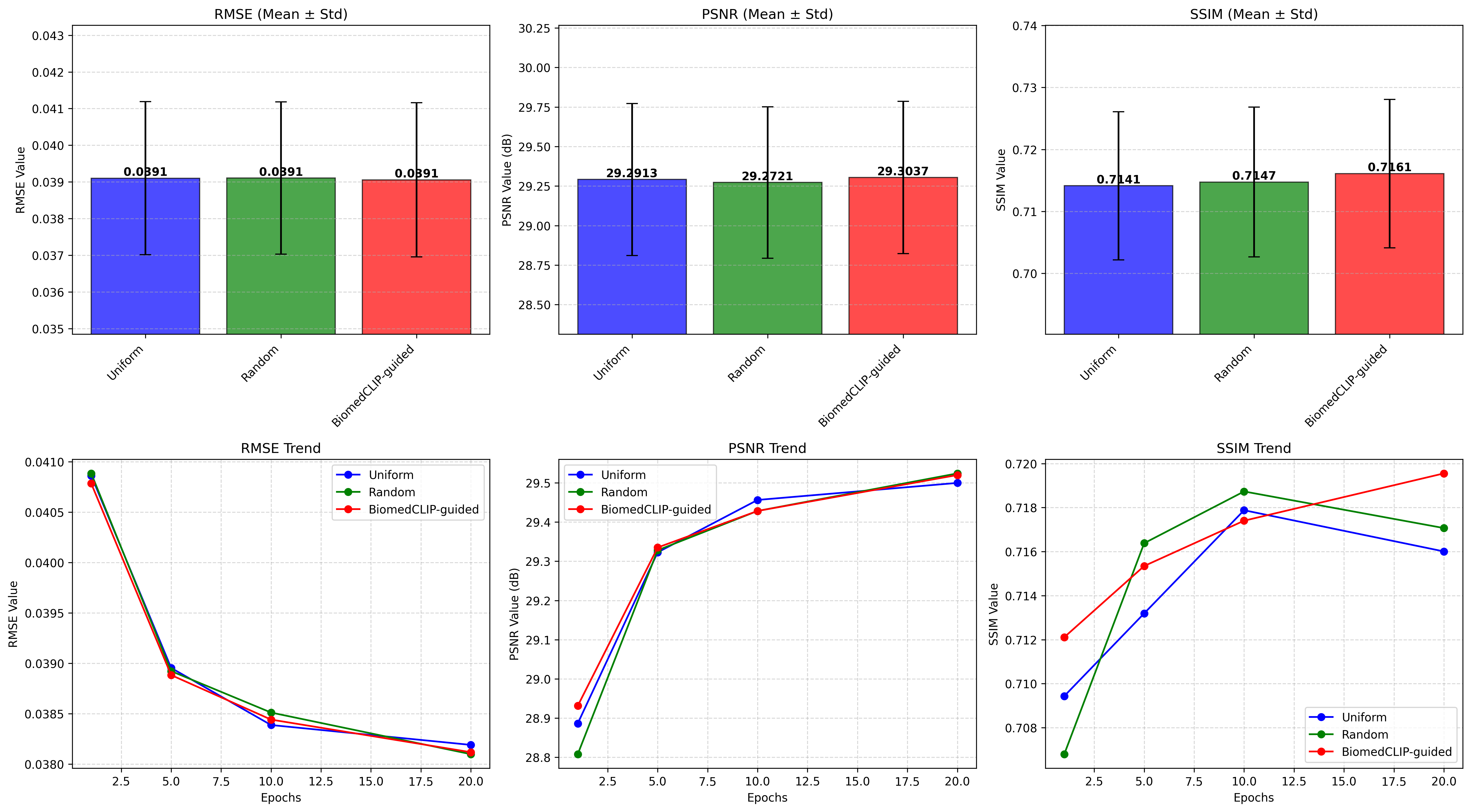}
    \captionof{figure}{Performance comparison of different attention weighting strategies. While RMSE and PSNR remain similar across all methods, BioAtt Block attention achieves the highest SSIM. The bottom row shows metric trends across training epochs. Notably, the BioAtt Block model exhibits a consistent and stable improvement in SSIM.}
    \label{fig6}
\end{center}

Our proposed model extends the conventional spatial attention mechanism (CBAM~\cite{woo2018cbam}) by expanding the typical two-channel pooling descriptors (average and max) into a multi-channel structure guided by organ-specific features extracted from BiomedCLIP. This modification introduces additional parameters, but only within the attention module.

To verify that the performance gain is not simply due to increased parameter count, we compare our BioAtt Block attention strategy with two alternative weighting methods: (1) Uniform Weighting: Equal weights are assigned to all organs. (\(\frac{1}{17}\) if \(N = 17\)). (2) Random weighting: Organ weights are initialized randomly but normalized to the sum of $1$. 

The experimental results (Figure~\ref{fig6}) indicate that all three approaches achieve similar RMSE and PSNR values. However, BioAtt Block attention consistently yields higher SSIM, particularly as training progresses. The bottom row illustrates that while uniform and random weightings plateau or fluctuate in SSIM, the BioAtt Block model demonstrates a stable upward trend. These findings confirm that the performance improvements are attributable to anatomically meaningful guidance from BiomedCLIP rather than to increased complexity. The guided attention mechanism helps preserve structural integrity, as reflected in the improved SSIM over time.

\section{Results}

\begin{table}[htbp]
\centering
\renewcommand{\arraystretch}{1.2} 
\begin{tabular}{lccc}
\hline
\textbf{Methods} & \textbf{RMSE} & \textbf{PSNR} & \textbf{SSIM}\\
\hline
Base & $0.0404_{\scriptscriptstyle \pm 0.0042}$ & $28.89_{\scriptscriptstyle \pm 0.95}$ & $0.7107_{\scriptscriptstyle \pm 0.0232}$ \\
Channel & $\textbf{0.0389}_{\scriptscriptstyle \pm 0.0041}$ & $\textbf{29.33}_{\scriptscriptstyle \pm 0.95}$ & $0.7137_{\scriptscriptstyle \pm 0.0245}$ \\
Spatial & $0.0390_{\scriptscriptstyle \pm 0.0042}$ & $29.31_{\scriptscriptstyle \pm 0.96}$ & $0.7135_{\scriptscriptstyle \pm 0.0245}$ \\
Whole & $0.0405_{\scriptscriptstyle \pm 0.0042}$ & $27.95_{\scriptscriptstyle \pm 0.87}$ & $0.7127_{\scriptscriptstyle \pm 0.0223}$ \\
Uniform & $0.0391_{\scriptscriptstyle \pm 0.0042}$ & $29.29_{\scriptscriptstyle \pm 0.96}$ & $0.7141_{\scriptscriptstyle \pm 0.0239}$ \\
Random & $0.0391_{\scriptscriptstyle \pm 0.0042}$ & $29.27_{\scriptscriptstyle \pm 0.96}$ & $0.7147_{\scriptscriptstyle \pm 0.0242}$ \\
\hline
BioAtt & $\textbf{0.0391}_{\scriptscriptstyle \pm 0.0042}$ & $\textbf{29.30}_{\scriptscriptstyle \pm 0.96}$ & $\textbf{0.7161}_{\scriptscriptstyle \pm 0.0239}$ \\
\hline
\end{tabular}
\vspace{1.0em}
\caption{Quantitative comparison of different models on LDCT denoising using RMSE, PSNR, and SSIM (mean $\pm$ standard deviation). The proposed \textit{BioAtt} model achieves the best SSIM while maintaining competitive RMSE and PSNR values.}
\label{tab:evaluation}
\end{table}

Table~\ref{tab:evaluation} presents a quantitative comparison of various models using three standard evaluation metrics: RMSE, PSNR, and SSIM. Among all variants, our proposed \textit{BioAtt} model achieves the highest SSIM ($0.7161$) while maintaining competitive performance in RMSE ($0.0391$) and PSNR ($29.30$). Interestingly, although \textit{Channel} model slightly outperforms others in RMSE and PSNR, it falls behind \textit{BioAtt} in terms of structural preservation, as indicated by the SSIM score. The \textit{Whole} image training strategy underperforms across all metrics which confirms the importance of localized feature learning via patch-based training. Furthermore, while \textit{Uniform} and \textit{Random} weighting yield comparable performance to \textit{BioAtt} in RMSE and PSNR, only \textit{BioAtt} consistently excels in SSIM metrics.

\begin{center}
\includegraphics[width=0.4\textwidth]{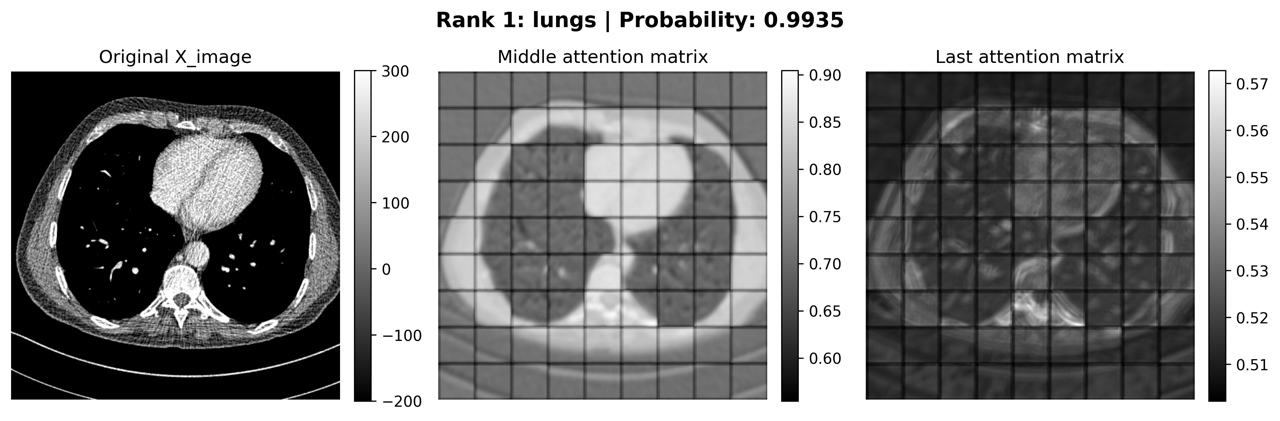}
\includegraphics[width=0.4\textwidth]{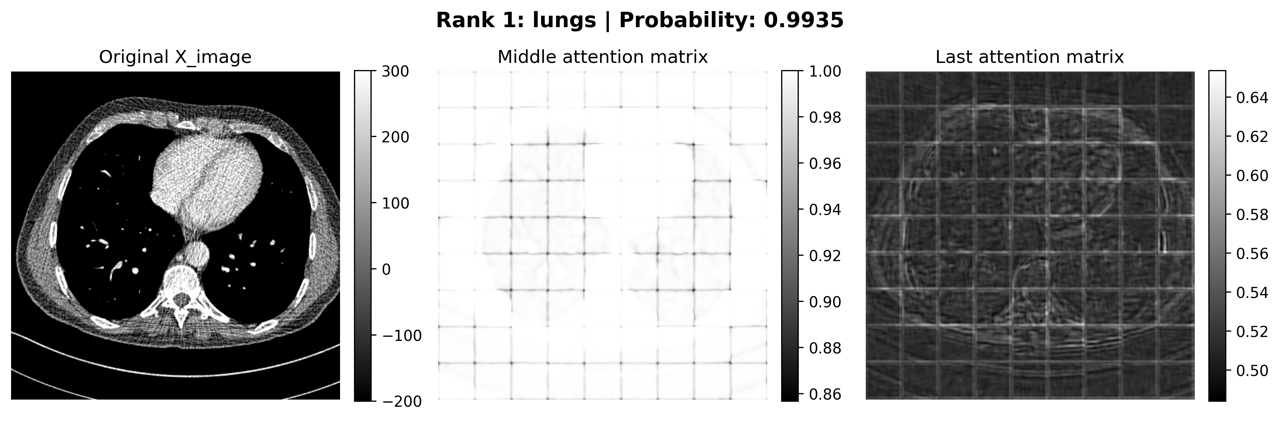}
\includegraphics[width=0.4\textwidth]{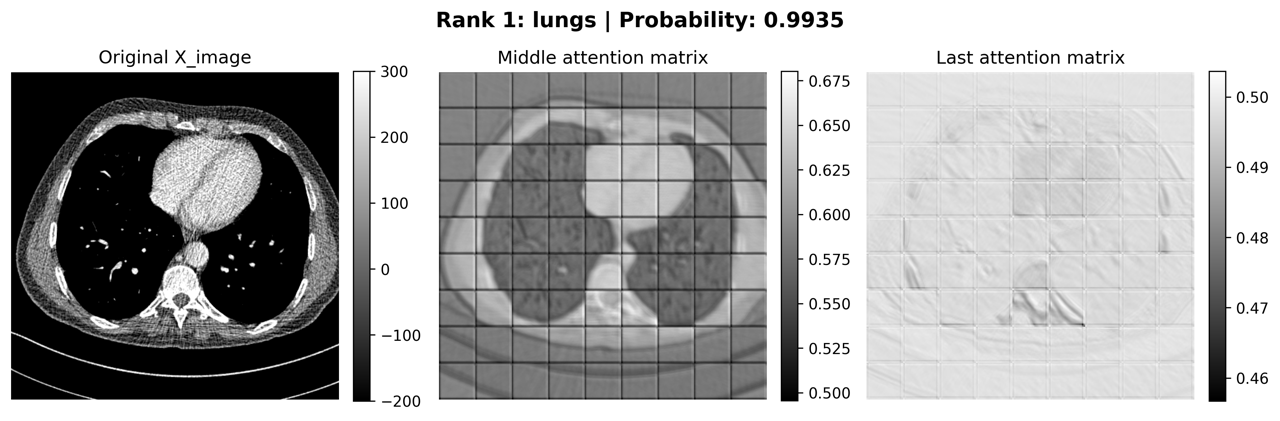}
\includegraphics[width=0.4\textwidth]{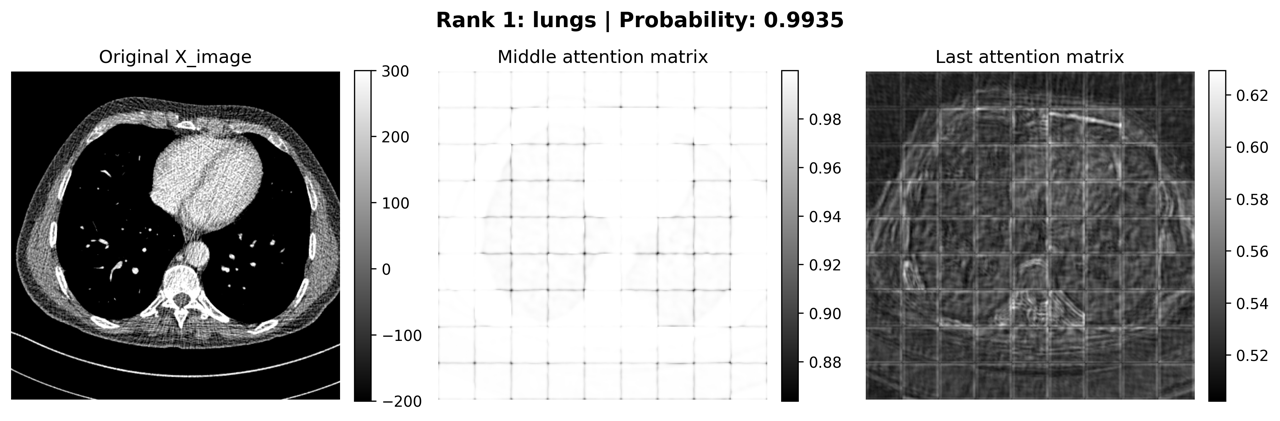}
\includegraphics[width=0.4\textwidth]{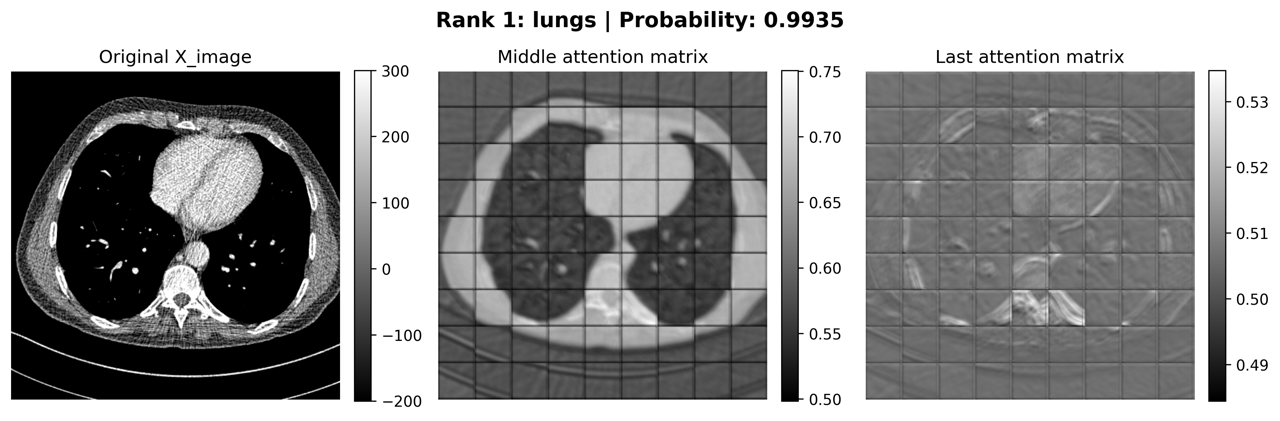}
\includegraphics[width=0.4\textwidth]{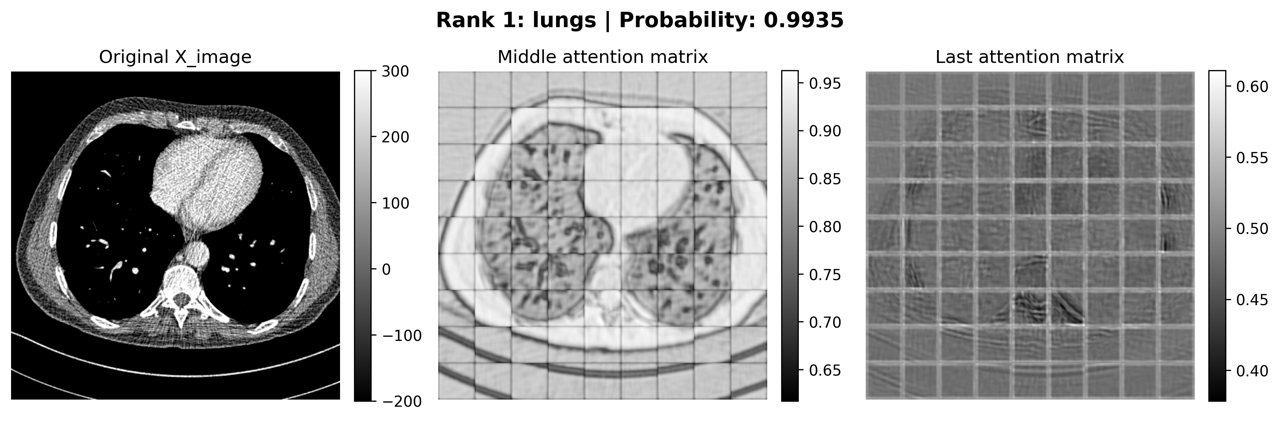}
\captionof{figure}{Visualization of attention maps: \textit{Random} (top row), \textit{Uniform} (middle row), and \textit{BioAtt} (bottom row) at two training stages: epoch 1 (left column) and epoch 20 (right column). All models are evaluated on the same lung CT image. \textit{BioAtt} already displays organ-specific attention with clear, contrast-reversed highlights at epoch 1. By epoch 20, both \textit{Random} and \textit{Uniform} models suffer from attention saturation. Meanwhile, \textit{BioAtt} maintains focused, anatomically meaningful attention which demonstrates its superior ability to preserve spatial and semantic relevance throughout training.}
\label{subsec:figc}
\end{center}

We compare the attention maps produced by three models: \textit{Random}, \textit{Uniform}, and our proposed \textit{BioAtt} model. The pretrained ViT transformer model assigns the test sample image the following prior probabilities: 0.9935 for lungs, 0.0059 for mediastinum, 0.0005 for spleen, and 0.0001 for ventricles, with near-zero values for all other organs.

At epoch 1 (Appendix~\ref{subsec:figb}), we observe that the attention map from the \textit{BioAtt} model already exhibits distinguishable patterns across organs. Although the grayscale distributions may appear similar, attention weights closer to $1$ are shown in white, while those closer to $0$ appear dark. This results in particularly noticeable contrast-reversed focal regions in the spine and liver, especially in the \textit{last} attention map.

In contrast, the \textit{Random} prior at epoch 1 (Appendix~\ref{subsec:figf}) results in an attention map that lacks variability among organs in both \textit{middle} and \textit{last} attention maps. This is because randomly assigned priors disrupt the network’s ability to learn spatially meaningful associations. On the other hand, the \textit{Uniform} prior at epoch 1 (Appendix~\ref{subsec:figd}) does exhibit some organ-level differentiation. However, compared to \textit{BioAtt}, the \textit{middle} attention maps appear washed out with narrower color value ranges. This suggests that uniform priors may under-emphasize dominant organs while overemphasizing irrelevant ones.

By epoch 20 (Appendix~\ref{subsec:figc}), a significant difference emerges between the models. In both \textit{Random} (Appendix~\ref{subsec:figg}) and \textit{Uniform} (Appendix~\ref{subsec:fige}) models, the \textit{middle} attention map becomes almost saturated with values close to 1, losing anatomical distinction and effectively neutralizing the role of the attention mechanism. This indicates that the attention module in these models is no longer functional. However, in \textit{BioAtt}, the \textit{middle} attention retains organ-aware variations, showing stronger differentiation between regions.

Interestingly, while the \textit{middle} attention fades in \textit{Random} and \textit{Uniform}, the \textit{last} attention maps in these models still retain some anatomical structure. This is because \textit{middle} attention is responsible for capturing lower-level, localized features, whereas \textit{last} attention integrates more global, high-level anatomical representations. As a result, even when the \textit{middle} attention is lost, the \textit{last} attention can still hold some residual organ-specific features.

Another key observation is that \textit{BioAtt} exhibits greater variation in its attention maps, particularly in the \textit{middle} layers. This increased variability is due to the anatomical priors incorporated via BiomedCLIP. When a specific organ is absent, BiomedCLIP assigns it a prior probability of zero, but this does not force the attention map itself to be entirely zero. As a result, the attention mechanism can still dynamically adapt, leading to larger variations in attention values. This characteristic is particularly evident in the \textit{middle} layers, where the attention maps display a wider range of intensity variations, ultimately contributing to a clearer and more distinct visual output.

In summary, while RMSE and PSNR values remain comparable across models, \textit{BioAtt} significantly outperforms other weighting mechanisms in terms of SSIM and long-term attention consistency. This demonstrates that BiomedCLIP-guided priors help models retain organ-specific focus throughout training, thereby reinforcing structural integrity and interpretability in LDCT denoising.

\section{Conclusion}

\textbf{BioAtt} effectively maintains competitive RMSE and PSNR performance while achieving consistent improvements in SSIM as training progresses. Unlike conventional spatial attention mechanisms, our BioAtt model preserves anatomically meaningful focus across training epochs. This produces distinct and semantically aligned attention maps that highlight anatomical structures.

Despite its strengths, the performance of BioAtt is influenced by the selection and diversity of text descriptions used for anatomical prior estimation. Expanding or refining the text inputs could further modify organ recognition and attention accuracy. Additionally, although BioAtt successfully differentiates regions of interest at a semantic level, the spatial distribution of attention maps remains visually similar across organs. To address this limitation, future work could integrate organ segmentation techniques—such as the Segment Anything Model (SAM) to generate explicit organ masks \cite{Zhu2024SAMAttAP, Ma2023SegmentAI, Wu2023MedicalSA, kirillov2023segany}. By incorporating spatial constraints or auxiliary supervision with a anatomical prior-guided loss function, we can further improve the alignment between anatomical priors and attention responses. Future research may also explore refining prior-guided attention mechanisms using more diverse and context-aware textual descriptions \cite{Song2022CLIPMA, Zhou2021ExtractFD, Rao2021DenseCLIPLD}. These improvements could further optimize LDCT denoising performance while preserving anatomical fidelity.

\bibliographystyle{unsrt}
\bibliography{references}  

\appendix
\begin{samepage}
\centering

\section{Spatial Attention Map}
\label{subsec:figa}
\includegraphics[width=0.5\textwidth]{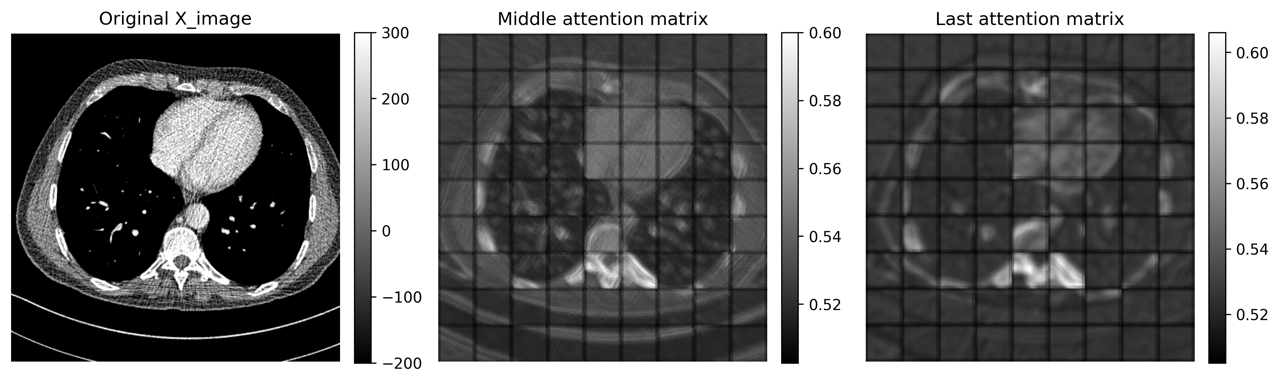}
\includegraphics[width=0.5\textwidth]{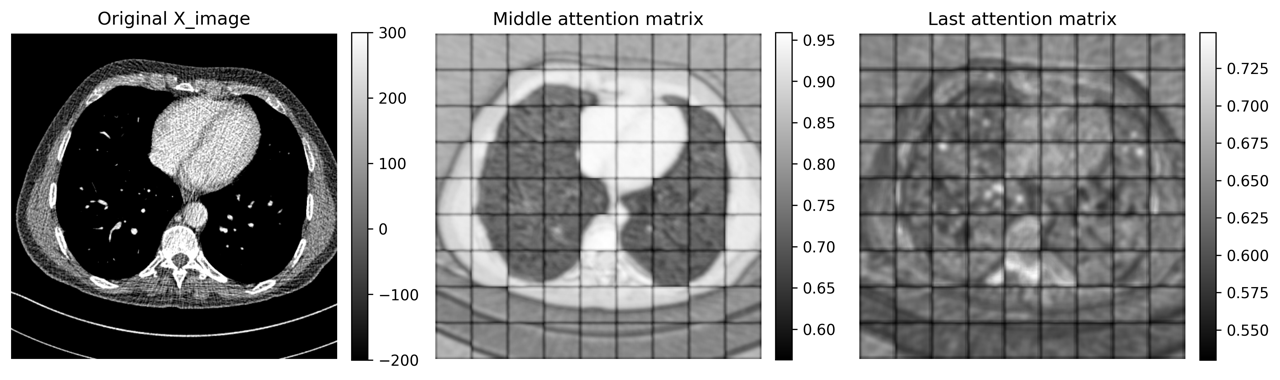}
\includegraphics[width=0.5\textwidth]{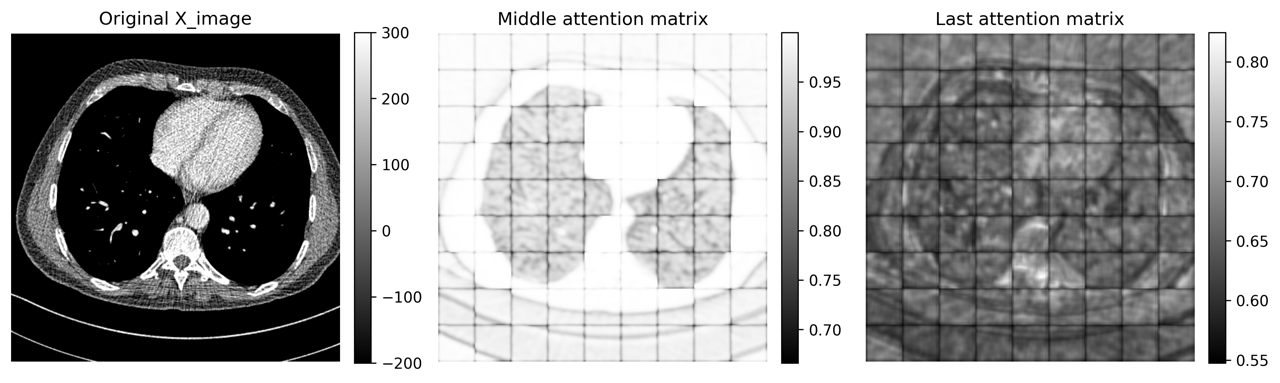}
\includegraphics[width=0.5\textwidth]{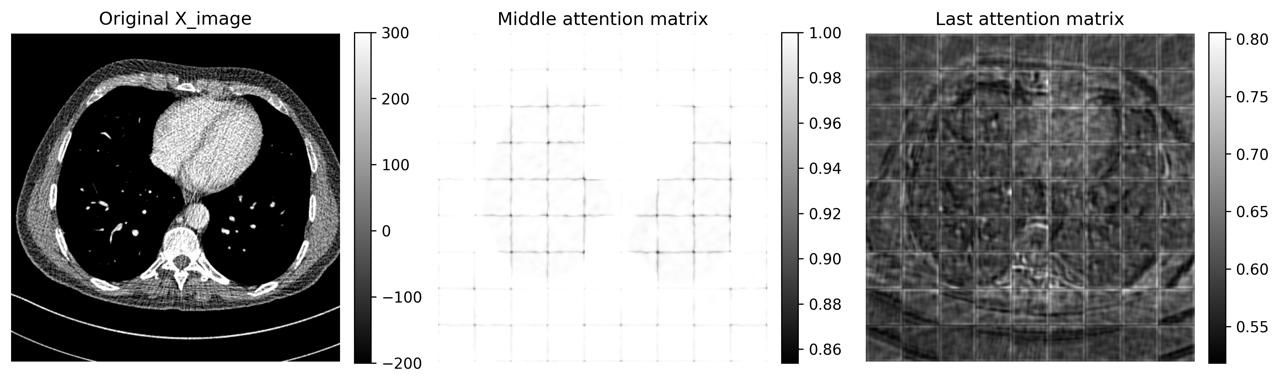}
\captionof{figure}{Intermediate Attention Map (Epoch 1, 5, 10, 20).}
\nopagebreak

\section{Attention Map with BioAtt Block: epoch=1}
\label{subsec:figb}
\includegraphics[width=0.4\textwidth]{bioatt_att_map/1/rank_1_lungs.png}
\includegraphics[width=0.4\textwidth]{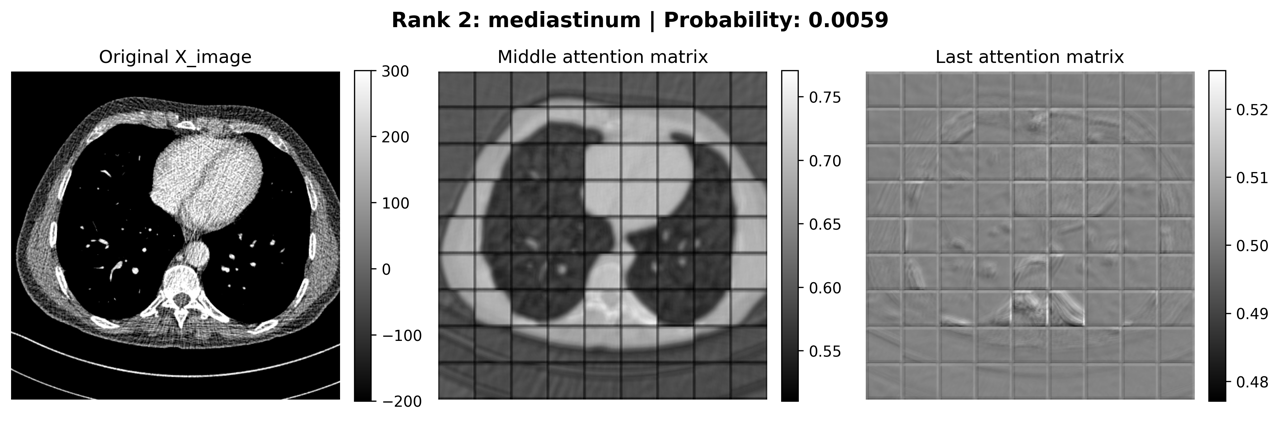}
\includegraphics[width=0.4\textwidth]{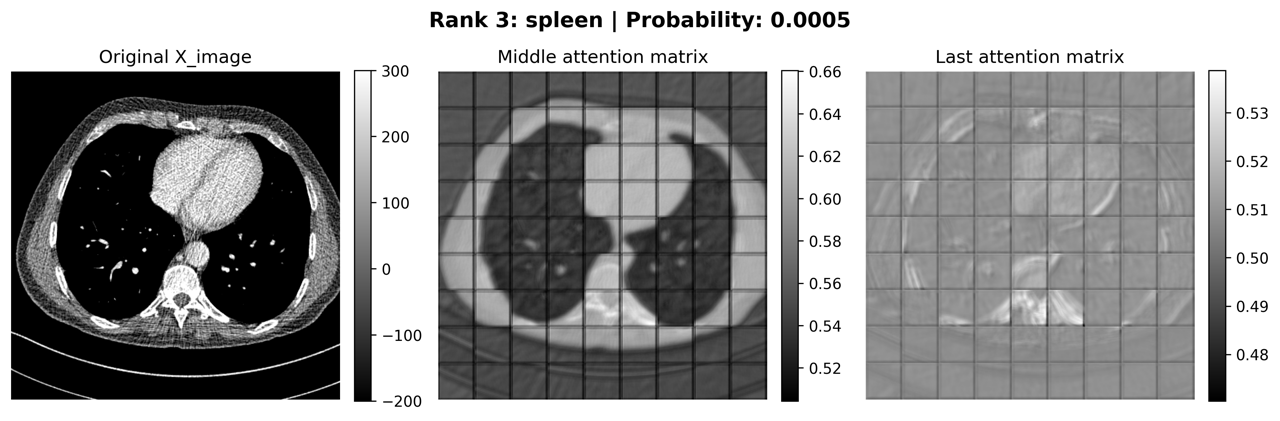}
\includegraphics[width=0.4\textwidth]{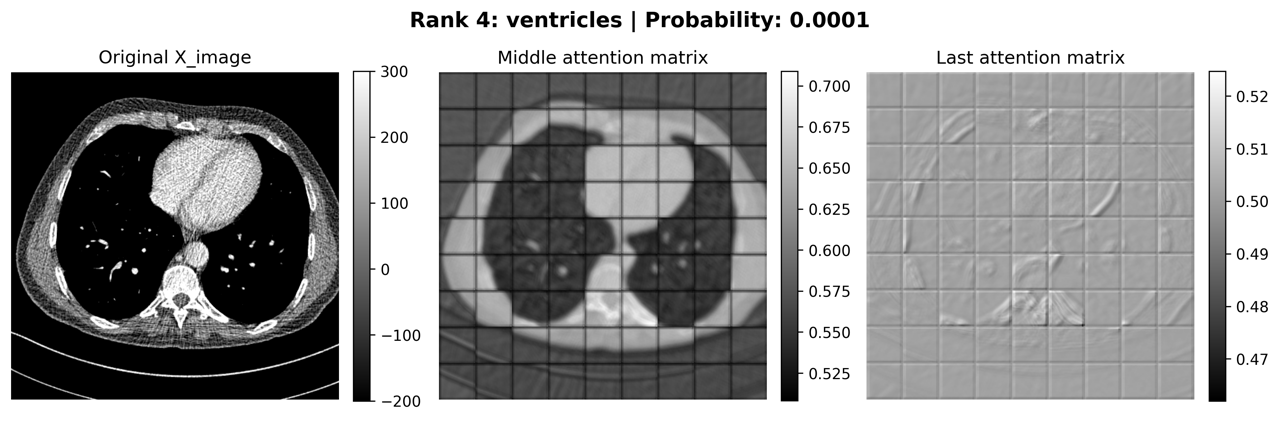}
\includegraphics[width=0.4\textwidth]{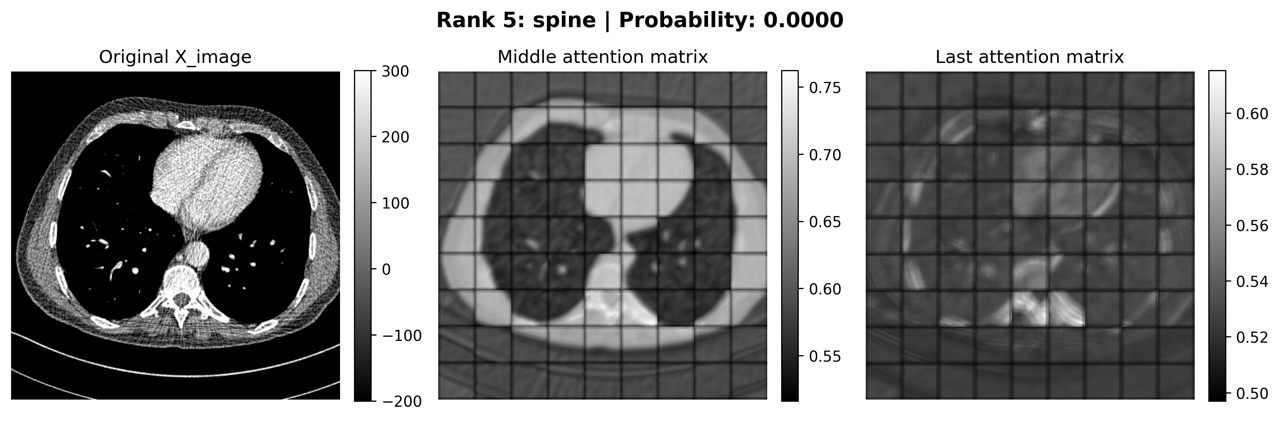}
\includegraphics[width=0.4\textwidth]{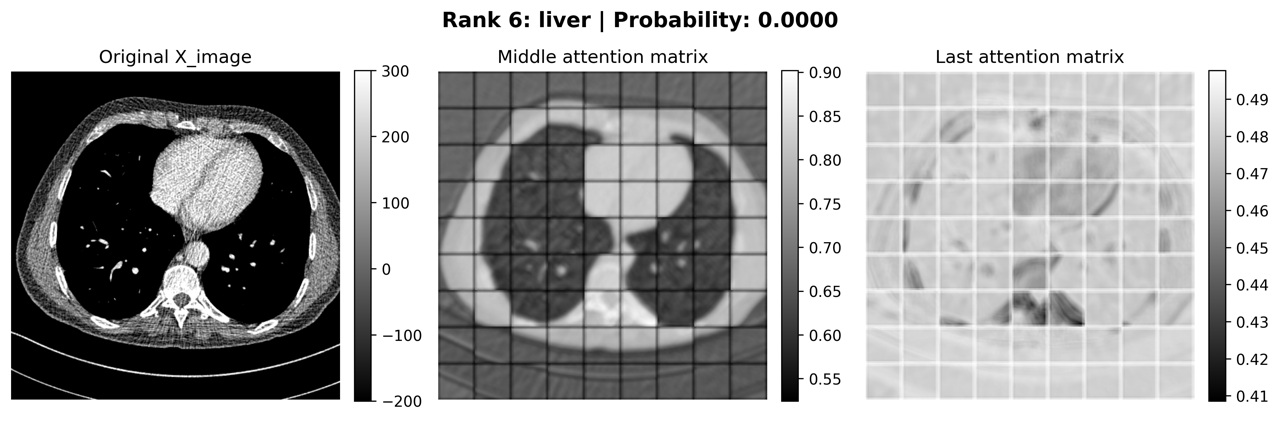}
\includegraphics[width=0.4\textwidth]{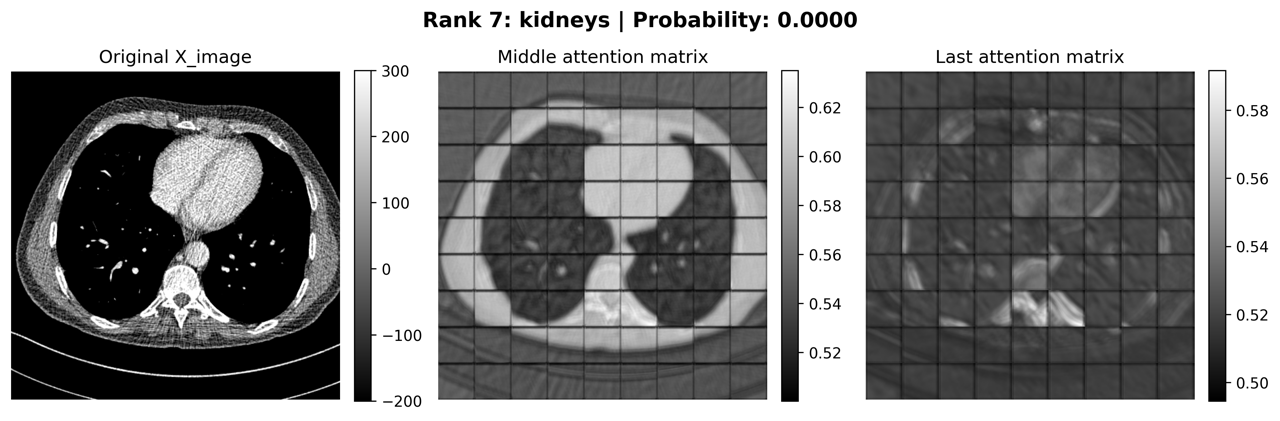}
\includegraphics[width=0.4\textwidth]{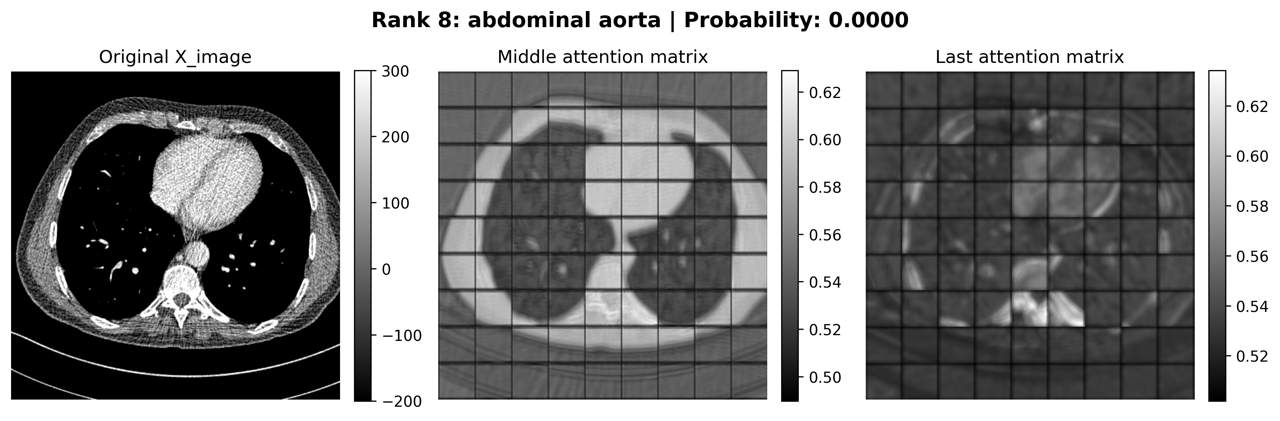}
\captionof{figure}{Intermediate Attention Map at Epoch 1.}
\nopagebreak

\section{Attention Map with BioAtt Block: epoch=20}
\label{subsec:figc}
\includegraphics[width=0.4\textwidth]{bioatt_att_map/20/rank_1_lungs.png}
\includegraphics[width=0.4\textwidth]{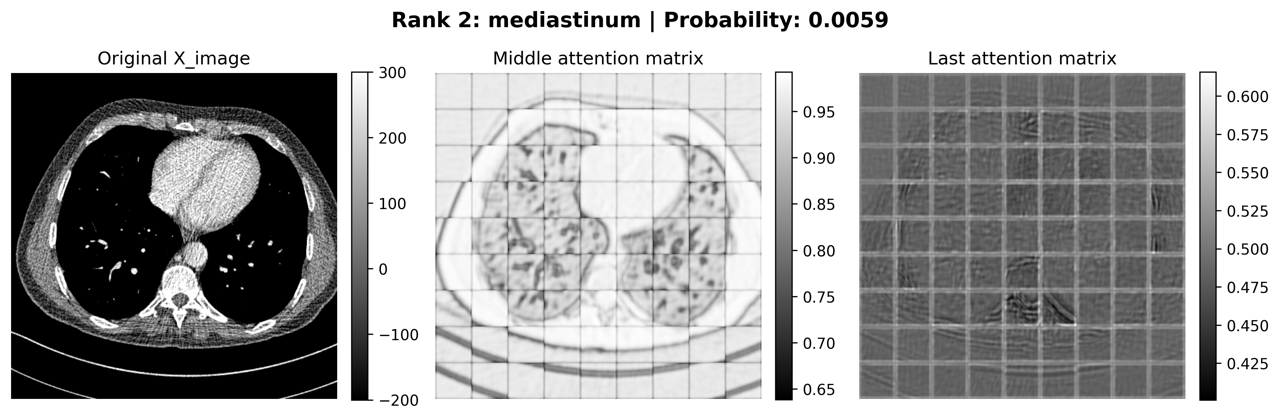}
\includegraphics[width=0.4\textwidth]{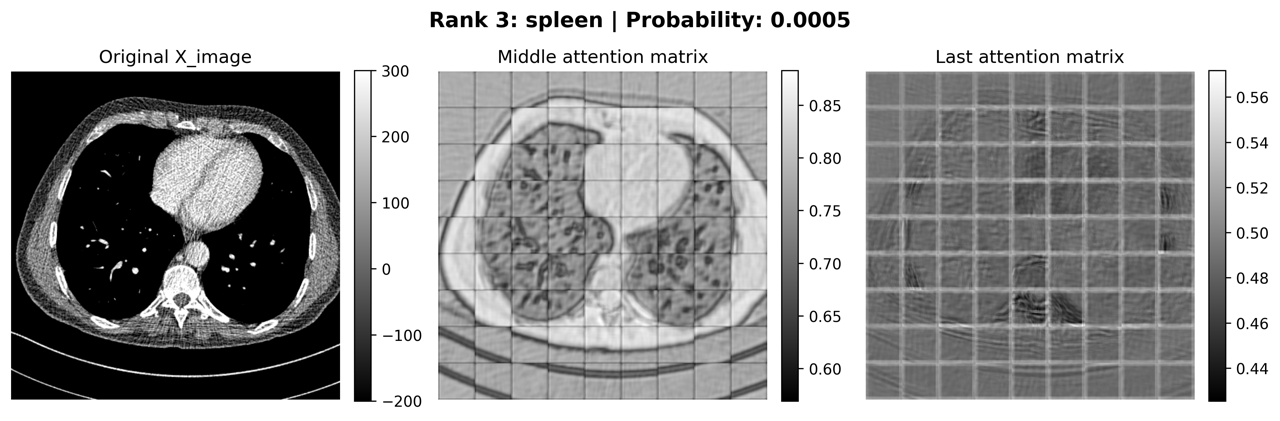}
\includegraphics[width=0.4\textwidth]{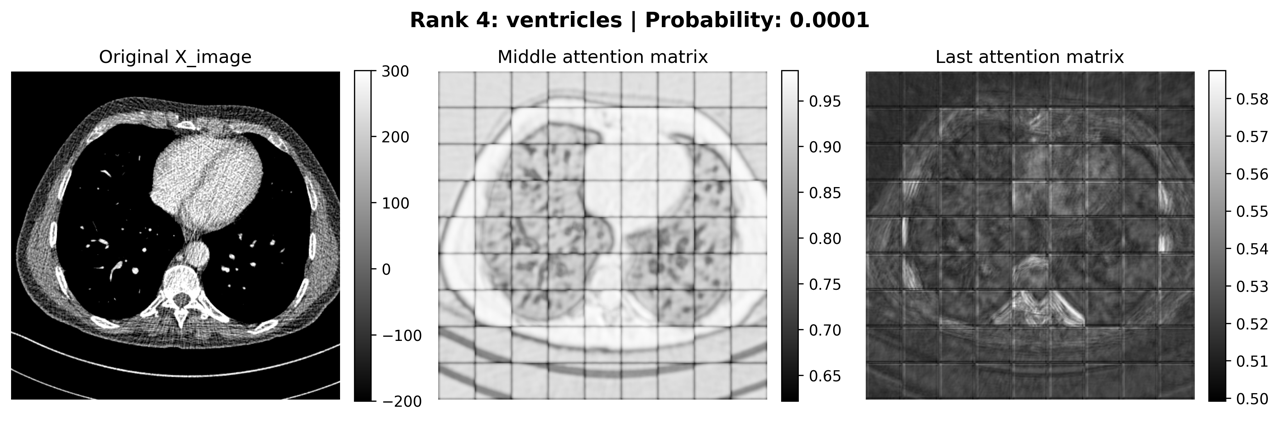}
\includegraphics[width=0.4\textwidth]{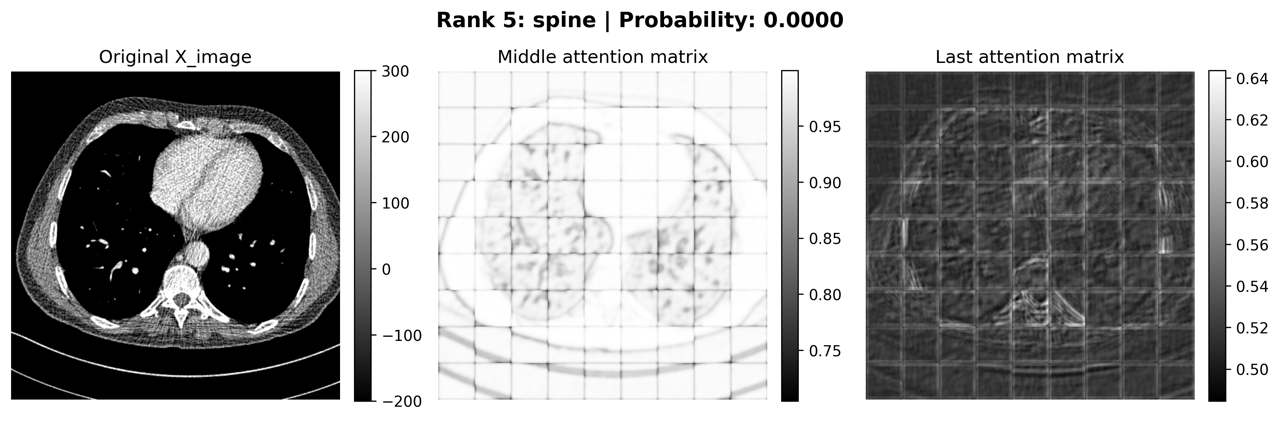}
\includegraphics[width=0.4\textwidth]{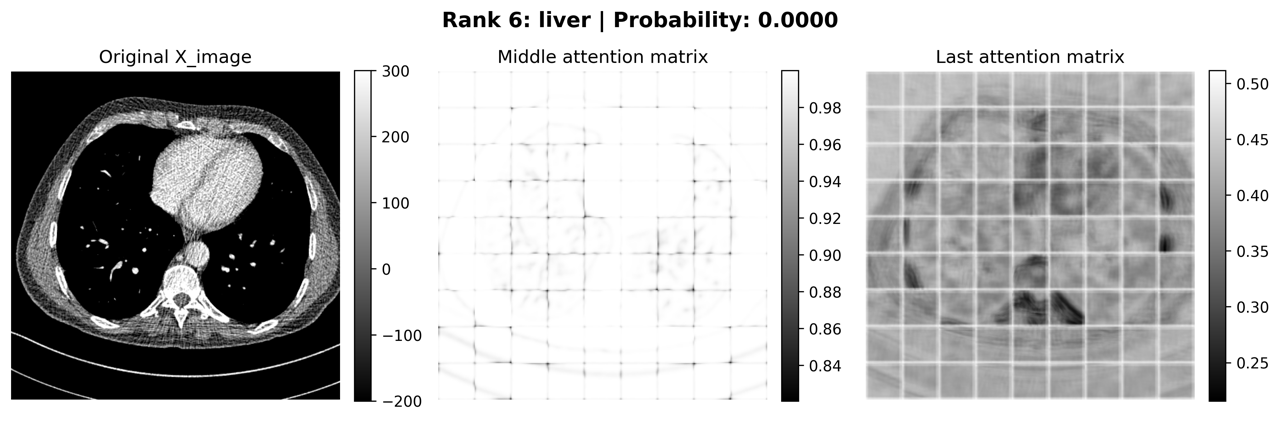}
\includegraphics[width=0.4\textwidth]{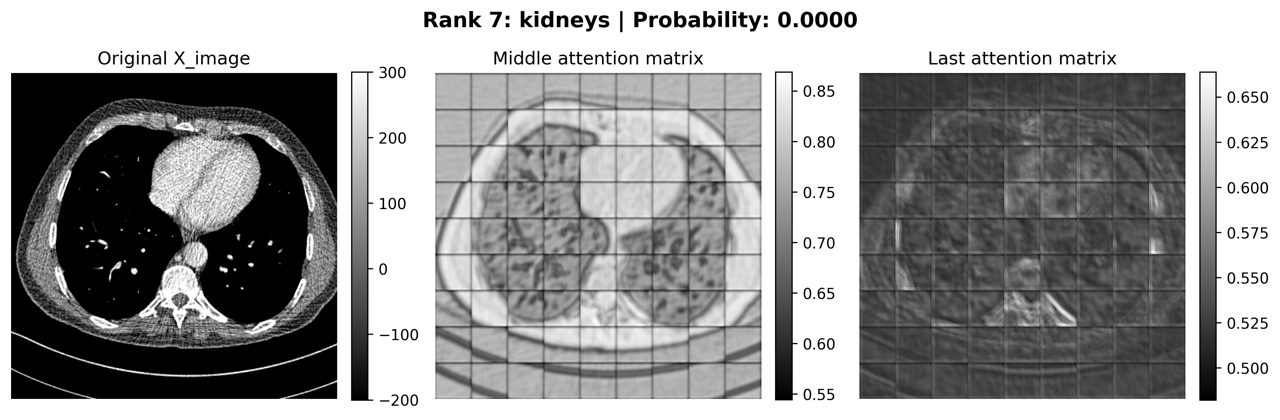}
\includegraphics[width=0.4\textwidth]{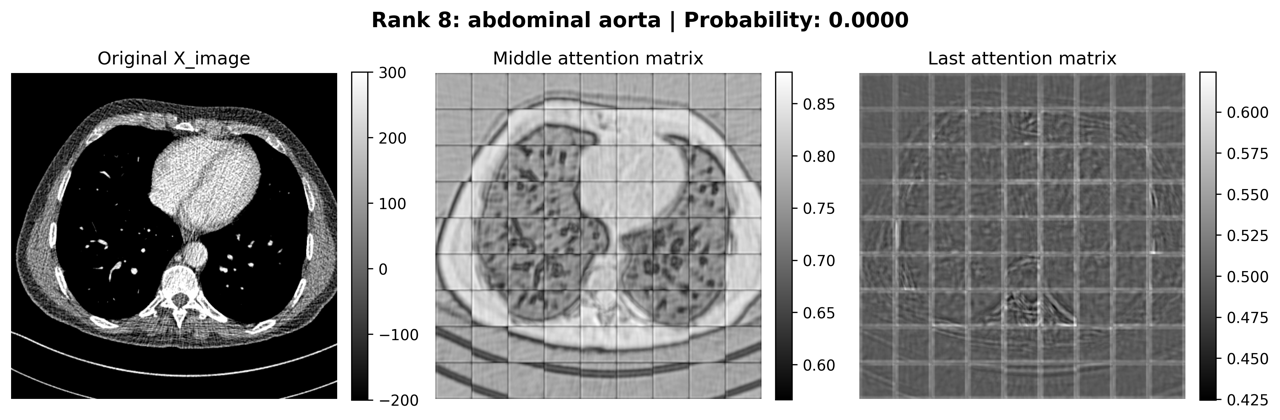}
\captionof{figure}{Intermediate Attention Map at Epoch 20.}
\nopagebreak

\section{Attention Map with Uniform Weighting: epoch=1}
\label{subsec:figd}
\includegraphics[width=0.4\textwidth]{uniform_att_map/1/rank_1_lungs.png}
\includegraphics[width=0.4\textwidth]{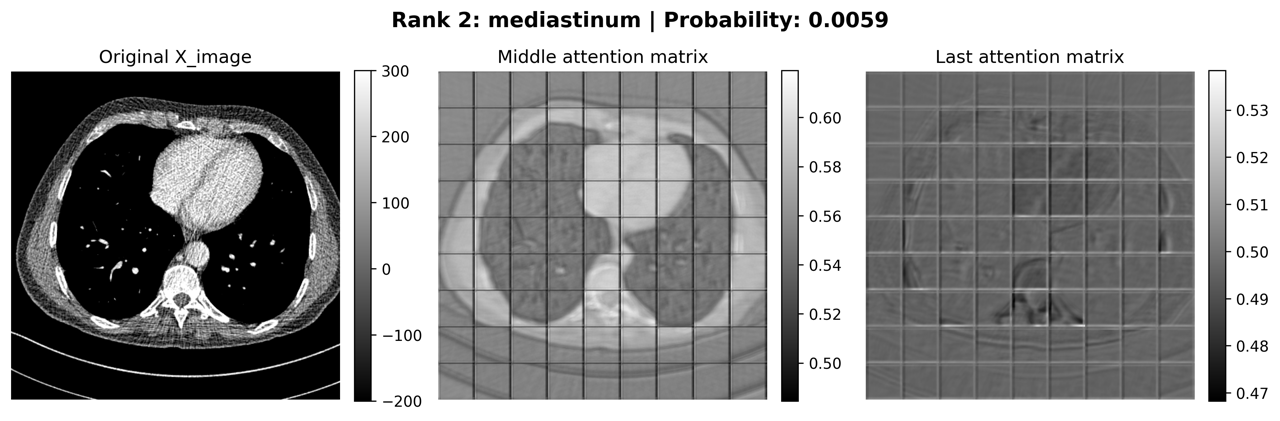}
\includegraphics[width=0.4\textwidth]{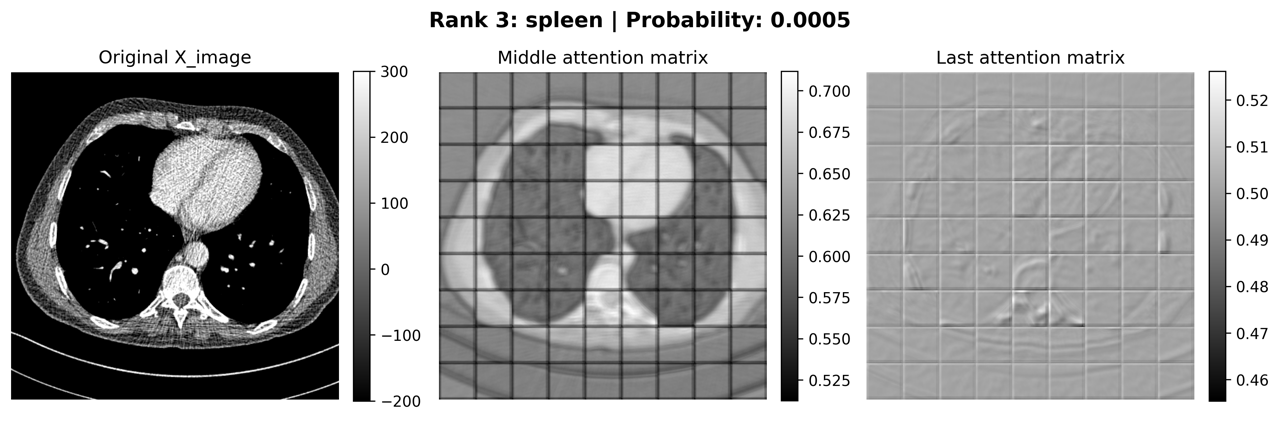}
\includegraphics[width=0.4\textwidth]{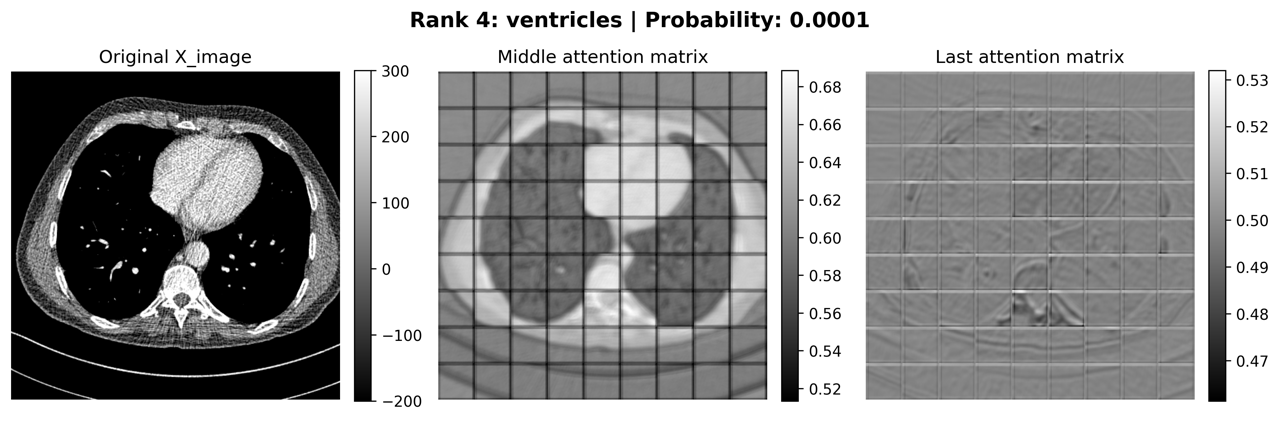}
\includegraphics[width=0.4\textwidth]{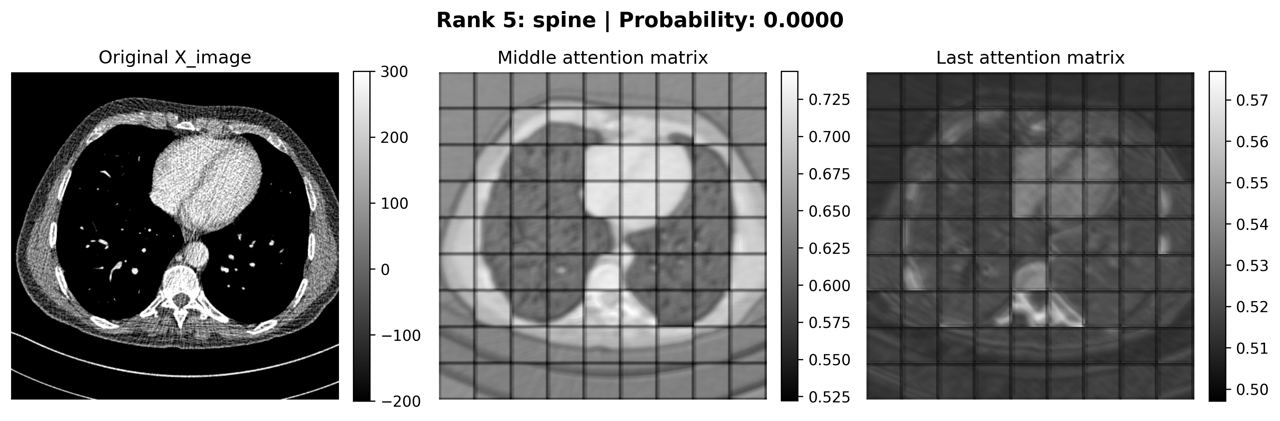}
\includegraphics[width=0.4\textwidth]{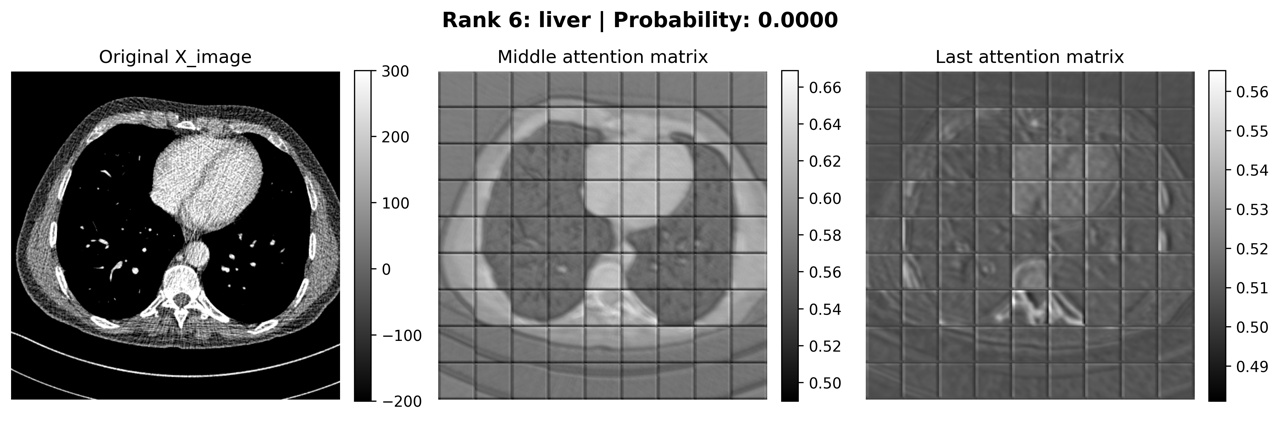}
\includegraphics[width=0.4\textwidth]{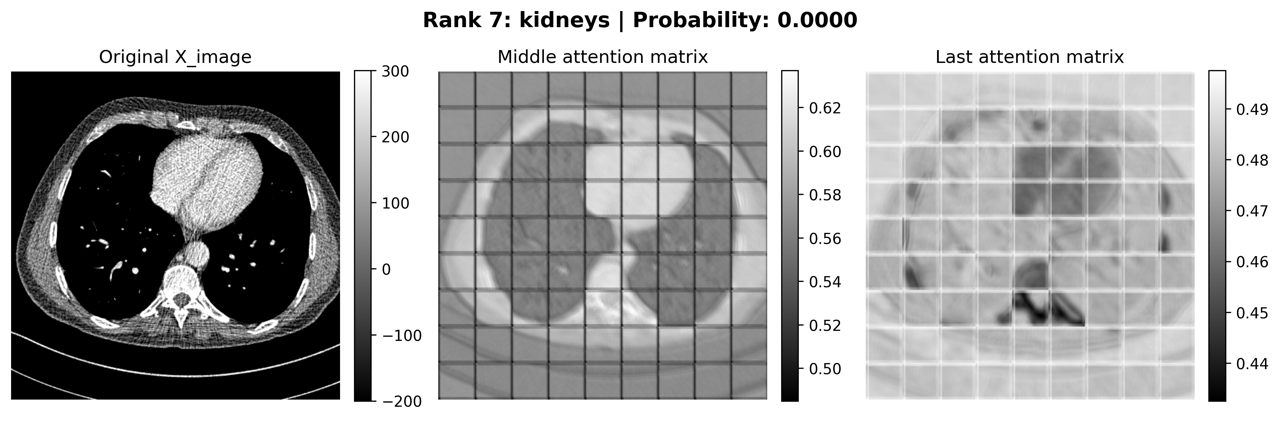}
\includegraphics[width=0.4\textwidth]{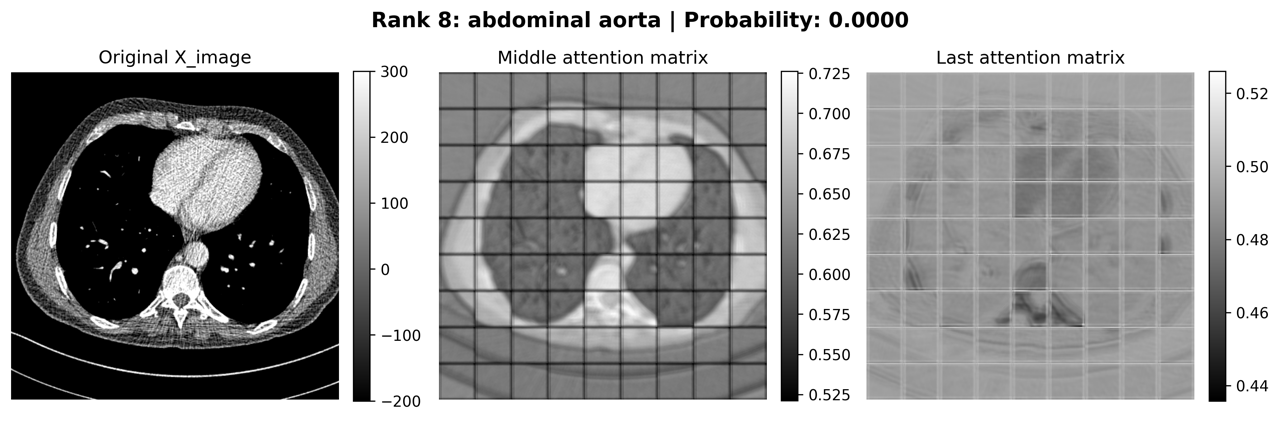}
\captionof{figure}{Intermediate Attention Map at Epoch 1.}
\nopagebreak

\section{Attention Map with Uniform Weighting: epoch=20}
\label{subsec:fige}
\includegraphics[width=0.4\textwidth]{uniform_att_map/20/rank_1_lungs.png}
\includegraphics[width=0.4\textwidth]{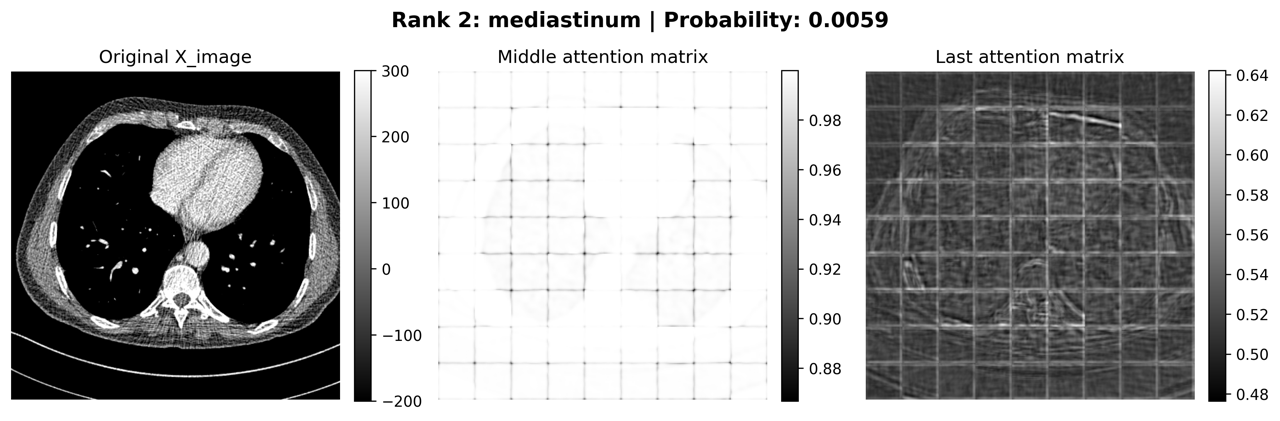}
\includegraphics[width=0.4\textwidth]{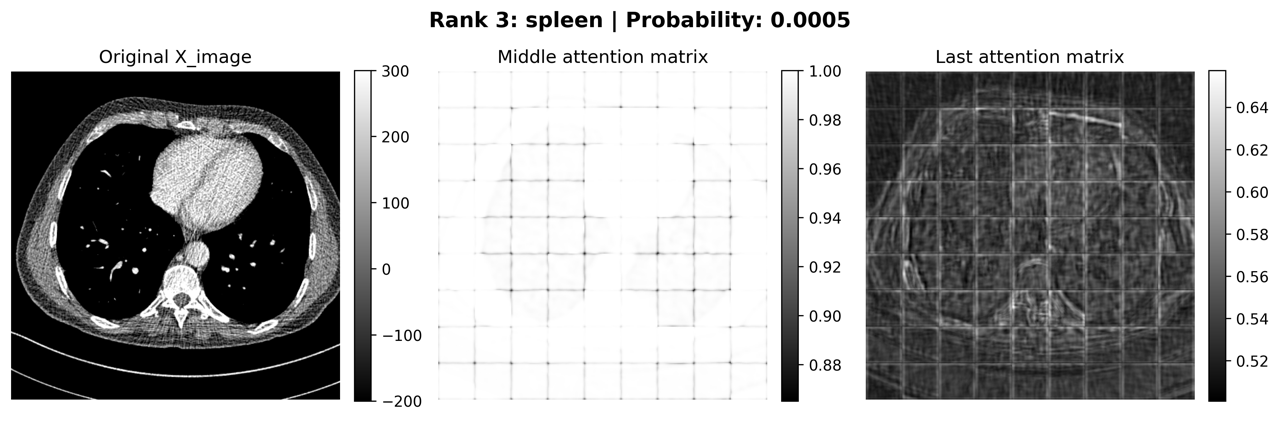}
\includegraphics[width=0.4\textwidth]{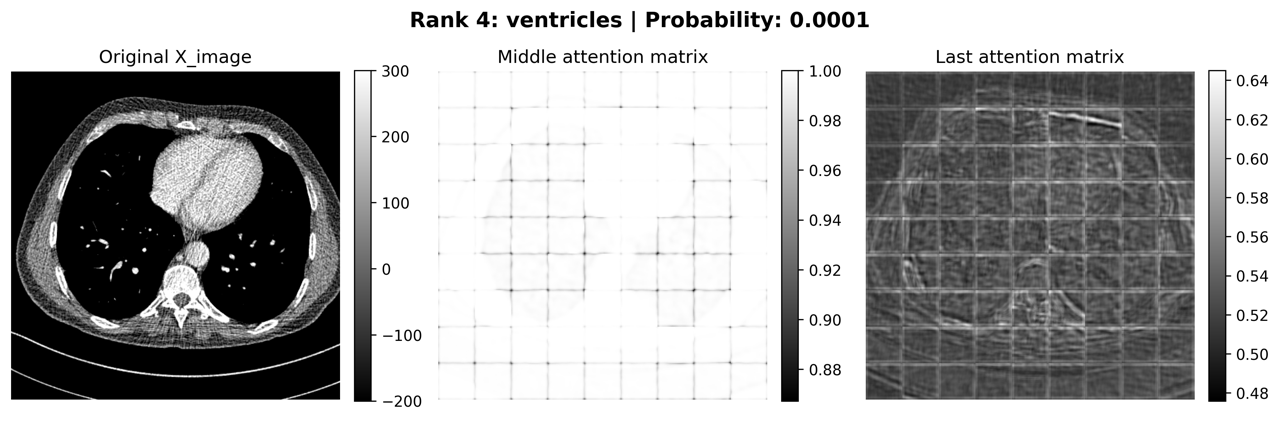}
\includegraphics[width=0.4\textwidth]{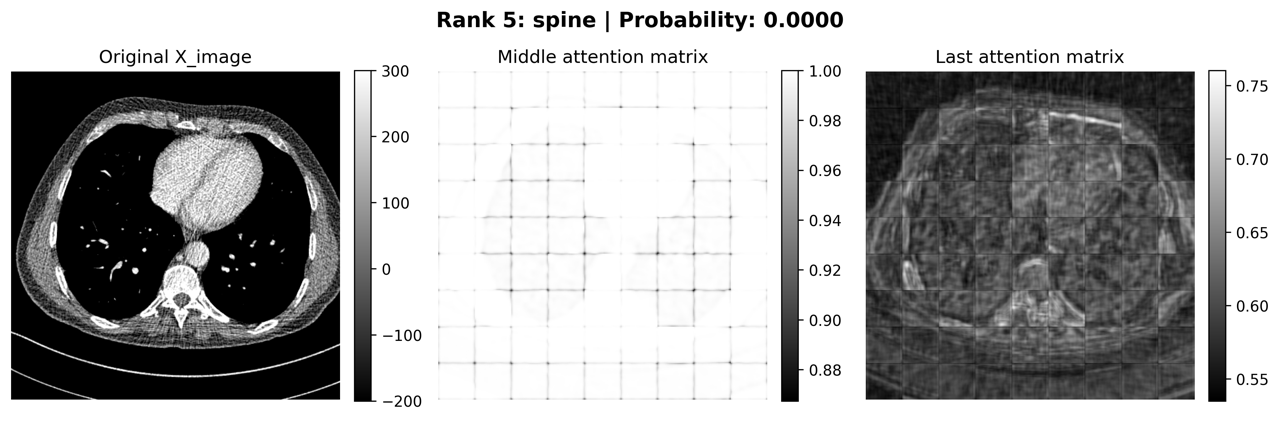}
\includegraphics[width=0.4\textwidth]{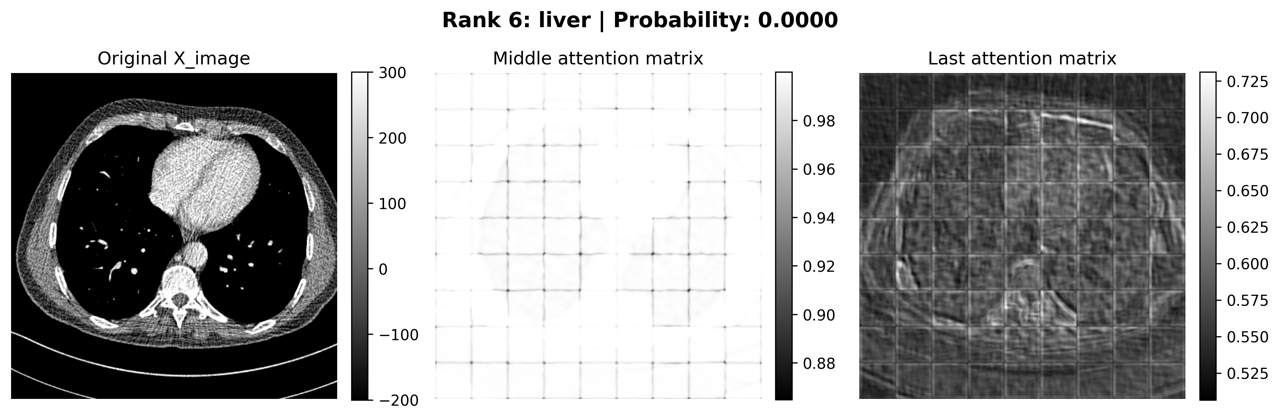}
\includegraphics[width=0.4\textwidth]{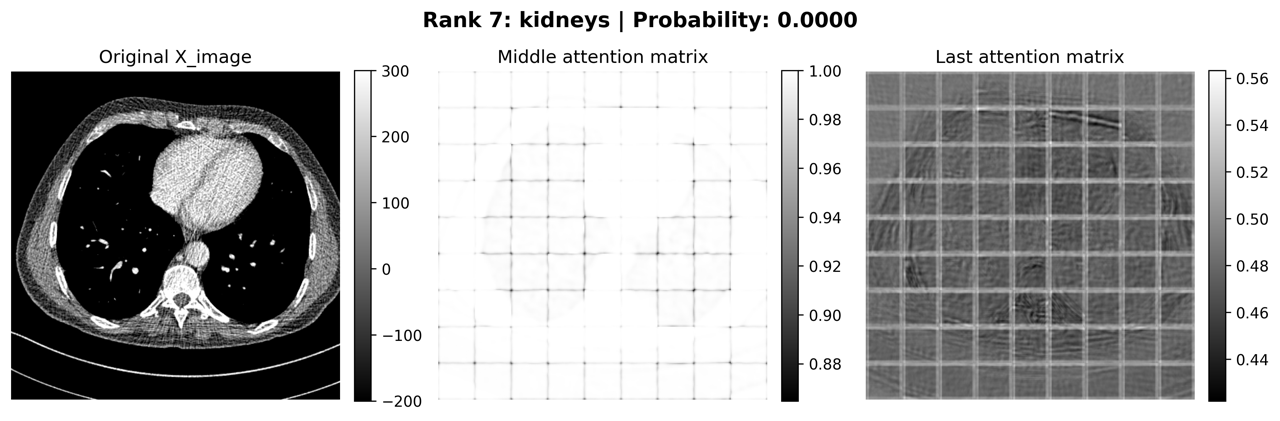}
\includegraphics[width=0.4\textwidth]{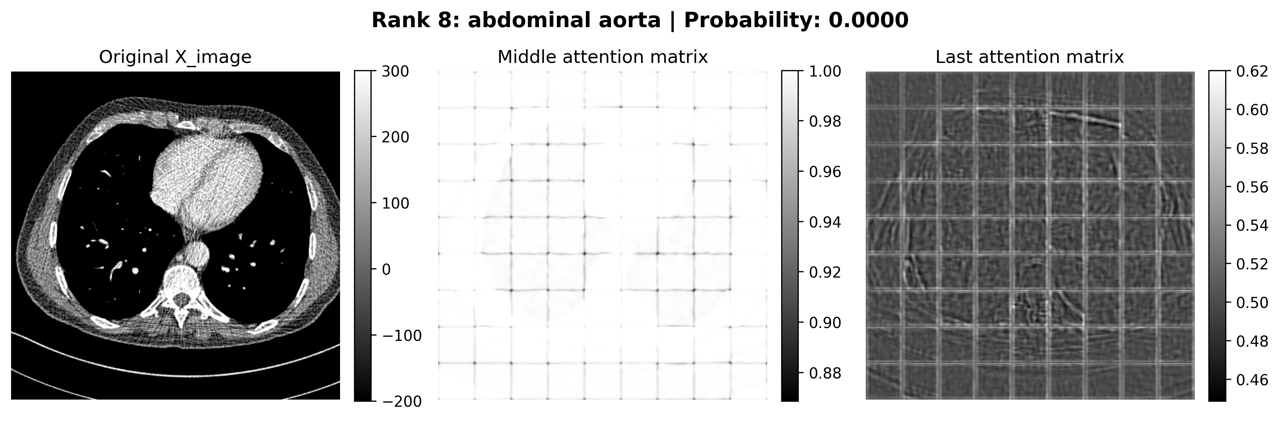}
\captionof{figure}{Intermediate Attention Map at Epoch 20.}
\nopagebreak

\section{Attention Map with Random Weighting: epoch=1}
\label{subsec:figf}
\includegraphics[width=0.4\textwidth]{random_att_map/1/rank_1_lungs.png}
\includegraphics[width=0.4\textwidth]{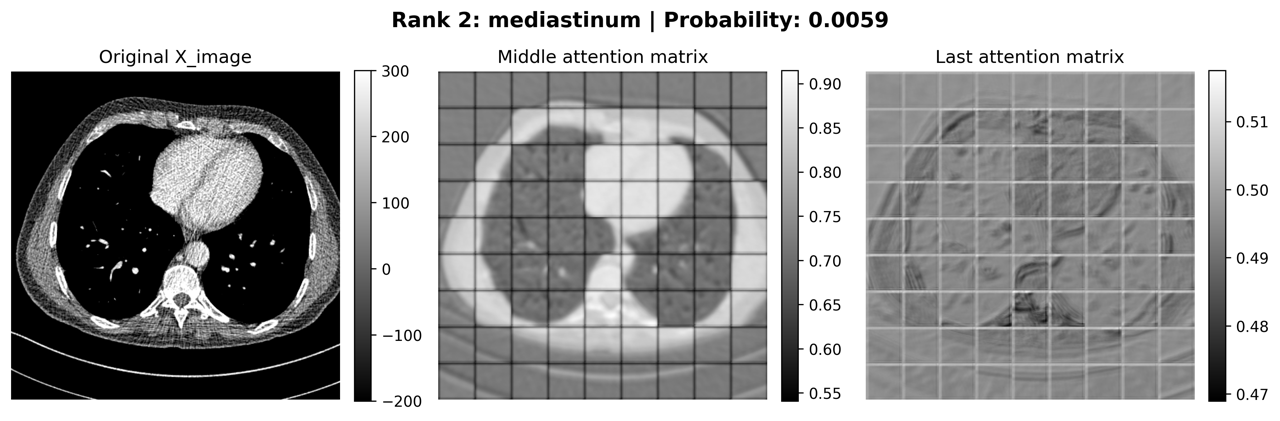}
\includegraphics[width=0.4\textwidth]{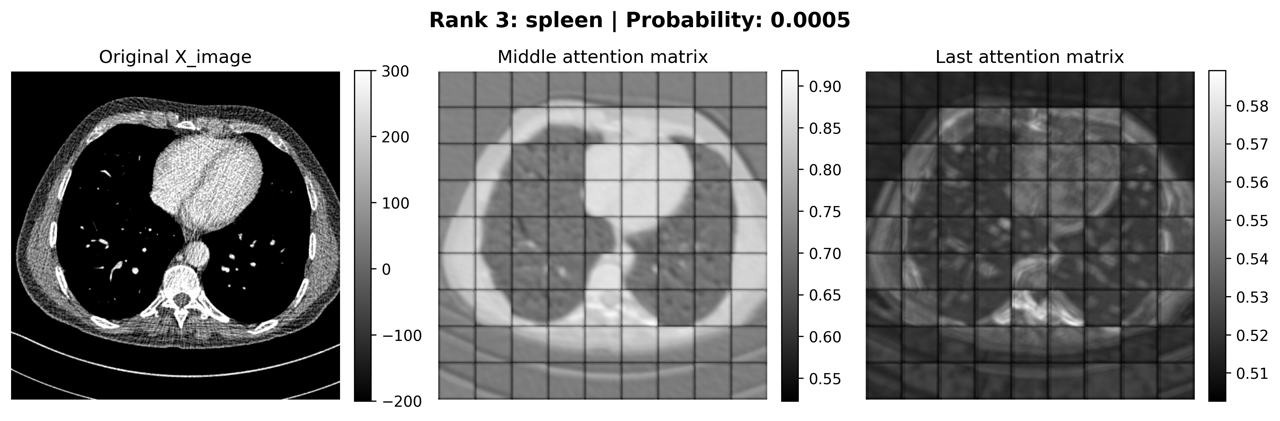}
\includegraphics[width=0.4\textwidth]{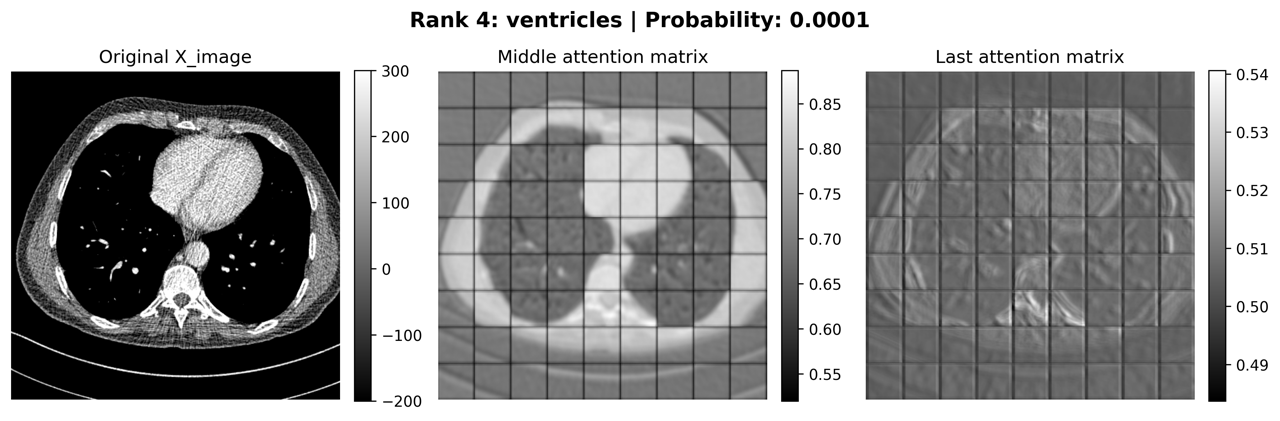}
\includegraphics[width=0.4\textwidth]{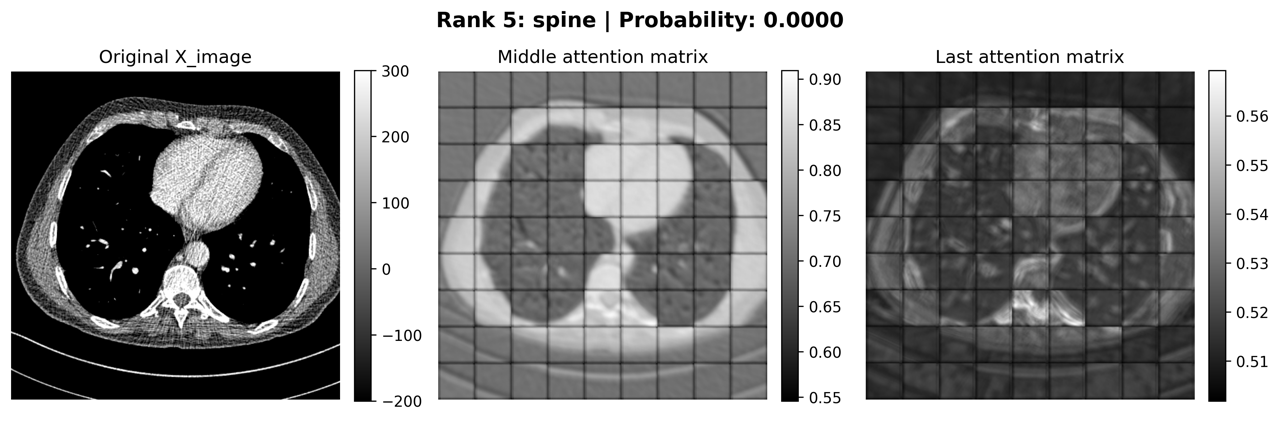}
\includegraphics[width=0.4\textwidth]{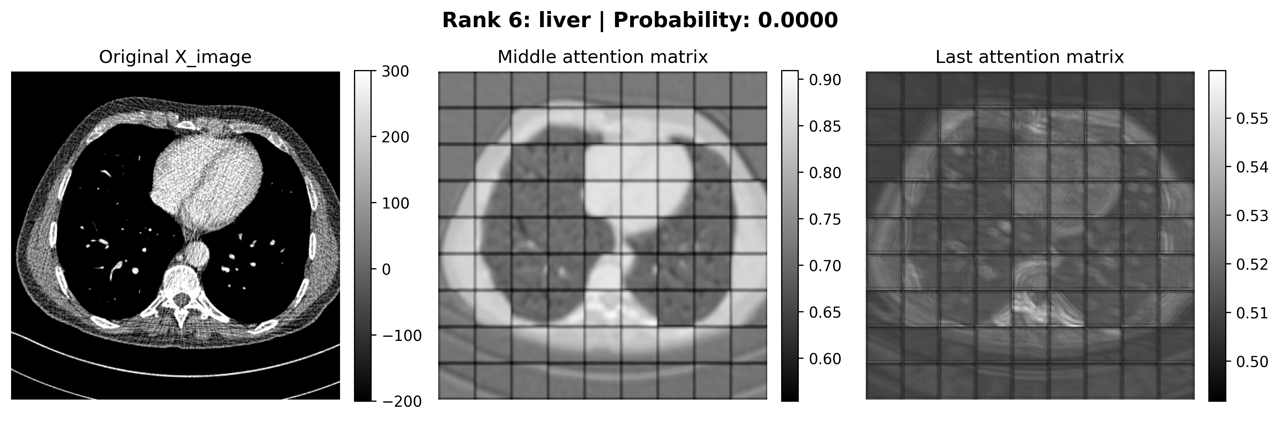}
\includegraphics[width=0.4\textwidth]{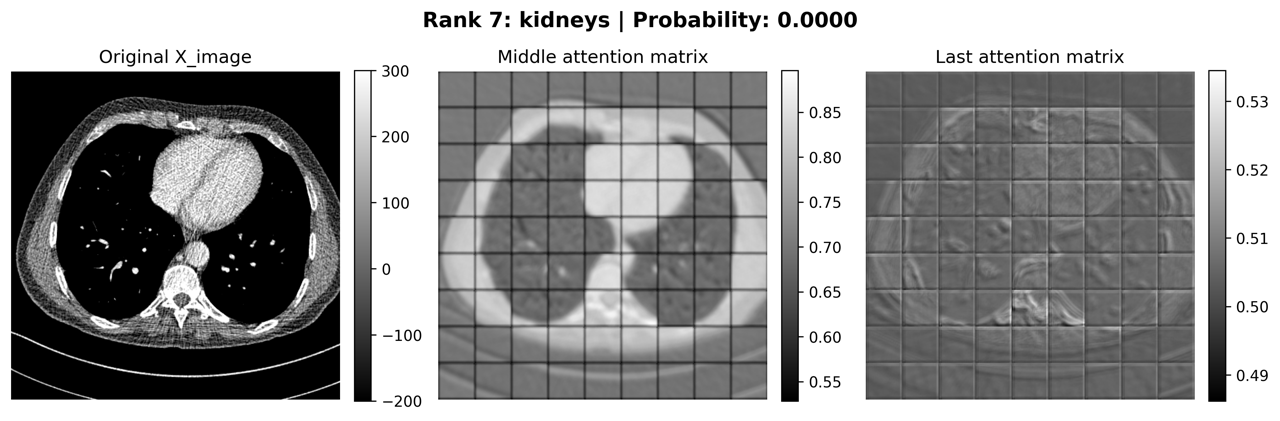}
\includegraphics[width=0.4\textwidth]{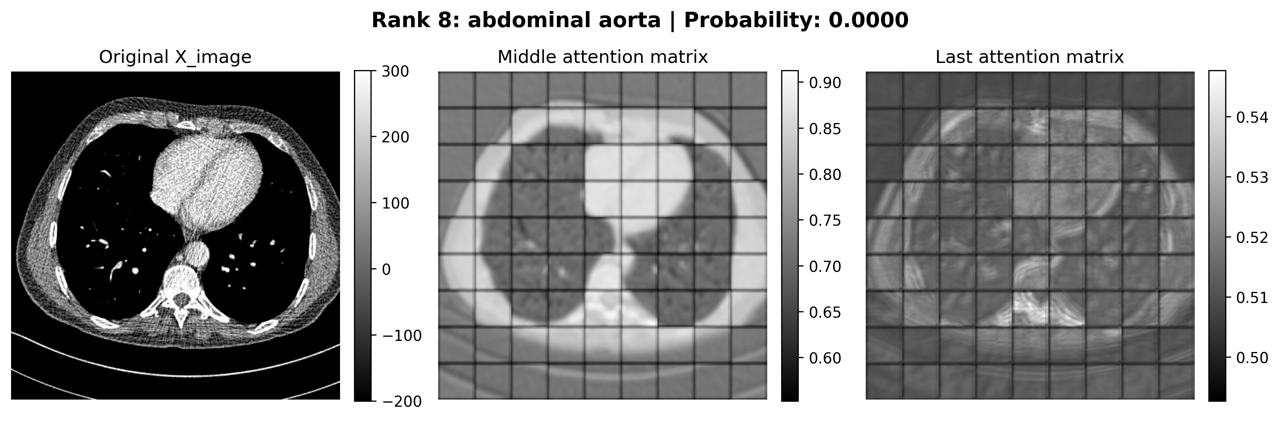}
\captionof{figure}{Intermediate Attention Map at Epoch 1.}
\nopagebreak

\section{Attention Map with Random Weighting: epoch=20}
\label{subsec:figg}
\includegraphics[width=0.4\textwidth]{uniform_att_map/20/rank_1_lungs.png}
\includegraphics[width=0.4\textwidth]{uniform_att_map/20/rank_2_mediastinum.png}
\includegraphics[width=0.4\textwidth]{uniform_att_map/20/rank_3_spleen.png}
\includegraphics[width=0.4\textwidth]{uniform_att_map/20/rank_4_ventricles.png}
\includegraphics[width=0.4\textwidth]{uniform_att_map/20/rank_5_spine.png}
\includegraphics[width=0.4\textwidth]{uniform_att_map/20/rank_6_liver.png}
\includegraphics[width=0.4\textwidth]{uniform_att_map/20/rank_7_kidneys.png}
\includegraphics[width=0.4\textwidth]{uniform_att_map/20/rank_8_abdominal_aorta.png}
\captionof{figure}{Intermediate Attention Map at Epoch 20.}
\nopagebreak

\end{samepage}

\end{document}